\documentclass[twocolumn,]{aastex631}
\usepackage{float}
\usepackage{longtable}
\usepackage{catchfile}

\shorttitle{X-ray and radio observations of Swift J1818.0–1607}
\shortauthors{Ibrahim et al.}

\graphicspath{{./}{figures/}}

\newcommand{\cxo}{{\em Chandra}}
\newcommand{\R}{{\em ROSAT}}
\newcommand{\A}{{\em ASCA}}

\newcommand{\e}{{\em Einstein}}
\newcommand{\xmm}{{\em XMM--Newton}}

\newcommand{\swift}{{\em Swift}}

\newcommand{\INT}{{\em INTEGRAL}}
\newcommand{\nustar}{{\em NuSTAR}}

\def\nh {$N_{\rm H}$}

\def\rchisq {$\chi_{\nu} ^{2}$}
\def\lum {erg\,s$^{-1}$}
\def\flux {erg\,s$^{-1}$cm$^{-2}$}
\def\ss {s\,s$^{-1}$}

\def\cm {cm$^{-2}$}
\def\arcsec{$^{\prime\prime}$}

\def\r{$R_{\rm BB}$}

\def\srclong{Swift\,J1818.0–1607}
\def\src{Swift\,J1818}

\begin{document}

\title{Deep X-ray and radio observations of the first outburst of the young magnetar Swift J1818.0–1607}

\correspondingauthor{A. Y. Ibrahim}
\email{ibrahim@ice.csic.es}

\author[0000-0002-5663-1712]{A. Y. Ibrahim}
\affiliation{Institute of Space Sciences (ICE, CSIC), Campus UAB, Carrer de Can Magrans s/n, E-08193, Barcelona, Spain}
\affiliation{Institut d’Estudis Espacials de Catalunya (IEEC), Carrer Gran Capità 2-4, E-08034 Barcelona, Spain}

\author[0000-0001-8785-5922]{A. Borghese}
\affiliation{Institute of Space Sciences (ICE, CSIC), Campus UAB, Carrer de Can Magrans s/n, E-08193, Barcelona, Spain}
\affiliation{Institut d’Estudis Espacials de Catalunya (IEEC), Carrer Gran Capità 2-4, E-08034 Barcelona, Spain}
\affiliation{Instituto de Astrofísica de Canarias, E-38205 La Laguna, Tenerife, Spain}
\affiliation{Departamento de Astrofísica, Universidad de La Laguna, E-38206 La Laguna, Tenerife, Spain}

\author[0000-0003-2177-6388]{N. Rea}
\affiliation{Institute of Space Sciences (ICE, CSIC), Campus UAB, Carrer de Can Magrans s/n, E-08193, Barcelona, Spain}
\affiliation{Institut d’Estudis Espacials de Catalunya (IEEC), Carrer Gran Capità 2-4, E-08034 Barcelona, Spain}

\author[0000-0001-7611-1581]{F. Coti Zelati}
\affiliation{Institute of Space Sciences (ICE, CSIC), Campus UAB, Carrer de Can Magrans s/n, E-08193, Barcelona, Spain}
\affiliation{Institut d’Estudis Espacials de Catalunya (IEEC), Carrer Gran Capità 2-4, E-08034 Barcelona, Spain}

\author[0000-0002-0430-6504]{E. Parent}
\affiliation{Institute of Space Sciences (ICE, CSIC), Campus UAB, Carrer de Can Magrans s/n, E-08193, Barcelona, Spain}
\affiliation{Institut d’Estudis Espacials de Catalunya (IEEC), Carrer Gran Capità 2-4, E-08034 Barcelona, Spain}

\author[0000-0002-7930-2276]{T. D. Russell}
\affiliation{INAF, Istituto di Astrofisica Spaziale e Fisica Cosmica, Via U. La Malfa 153, I-90146 Palermo, Italy}

\author[0000-0001-5116-6789]{S. Ascenzi}
\affiliation{Institute of Space Sciences (ICE, CSIC), Campus UAB, Carrer de Can Magrans s/n, E-08193, Barcelona, Spain}
\affiliation{Institut d’Estudis Espacials de Catalunya (IEEC), Carrer Gran Capità 2-4, E-08034 Barcelona, Spain}

\author[0000-0002-5254-3969]{R. Sathyaprakash}
\affiliation{Institute of Space Sciences (ICE, CSIC), Campus UAB, Carrer de Can Magrans s/n, E-08193, Barcelona, Spain}
\affiliation{Institut d’Estudis Espacials de Catalunya (IEEC), Carrer Gran Capità 2-4, E-08034 Barcelona, Spain}

\author[0000-0001-9494-0981]{D.~G\"otz} 
\affiliation{AIM-CEA/DRF/Irfu/Département d’Astrophysique, CNRS, Université Paris-Saclay, Université de Paris Cité, Orme des Merisiers, F-91191 Gif-sur-Yvette, France}

\author[0000-0003-3259-7801]{S.~Mereghetti} 
\affiliation{INAF—Istituto di Astrofisica Spaziale e Fisica Cosmica di Milano, via A. Corti 12, I-20133 Milano, Italy}

\author[0000-0002-5711-9278]{M.~Topinka} 
\affiliation{INAF—Istituto di Astrofisica Spaziale e Fisica Cosmica di Milano, via A. Corti 12, I-20133 Milano, Italy}

\author[0000-0001-6641-5450]{M.~Rigoselli} 
\affiliation{INAF—Istituto di Astrofisica Spaziale e Fisica Cosmica di Milano, via A. Corti 12, I-20133 Milano, Italy}

\author[0000-0001-6353-0808]{V.~Savchenko} 
\affiliation{Department of Astronomy, University of Geneva, Ch. d’Ecogia 16, 1290, Versoix, Switzerland}

\author[0000-0001-6278-1576]{S.~Campana} 
\affiliation{INAF--Osservatorio Astronomico di Brera, Via Bianchi 46, Merate (LC), I-23807, Italy}

\author[0000-0001-5480-6438]{G. L. Israel} 
\affiliation{INAF—Osservatorio Astronomico di Roma, via Frascati 33, I-00078 Monteporzio Catone, Italy}

\author[0000-0002-6038-1090]{A. Tiengo}
\affiliation{Scuola Universitaria Superiore IUSS Pavia, Palazzo del Broletto, piazza della Vittoria 15, I-27100 Pavia, Italy}
\affiliation{INAF—Istituto di Astrofisica Spaziale e Fisica Cosmica di Milano, via A. Corti 12, I-20133 Milano, Italy}
\affiliation{Istituto Nazionale di Fisica Nucleare (INFN), Sezione di Pavia, via A. Bassi 6, I-27100 Pavia, Italy}

\author[0000-0002-3635-5677]{R. Perna}
\affiliation{Department of Physics and Astronomy, Stony Brook University, Stony Brook, NY 11794, USA}
\affiliation{Center for Computational Astrophysics, Flatiron Institute, 162 5th Avenue, New York, NY 10010, USA}

\author[0000-0003-3977-8760]{R. Turolla} 
\affiliation{Dipartimento di Fisica e Astronomia “Galileo Galilei”, Università di Padova, via F. Marzolo 8, I-35131 Padova, Italy}
\affiliation{Mullard Space Science Laboratory, University College London, Holmbury St. Mary, Dorking, Surrey RH5 6NT, UK}

\author[0000-0001-5326-880X]{S. Zane} 
\affiliation{Mullard Space Science Laboratory, University College London, Holmbury St. Mary, Dorking, Surrey RH5 6NT, UK}

\author[0000-0003-4849-5092]{P. Esposito} 
\affiliation{Scuola Universitaria Superiore IUSS Pavia, Palazzo del Broletto, piazza della Vittoria 15, I-27100 Pavia, Italy}
\affiliation{INAF—Istituto di Astrofisica Spaziale e Fisica Cosmica di Milano, via A. Corti 12, I-20133 Milano, Italy}

\author[0000-0003-3952-7291]{G. A. Rodríguez Castillo} 
\affiliation{INAF, Istituto di Astrofisica Spaziale e Fisica Cosmica, Via U. La Malfa 153, I-90146 Palermo, Italy}

\author[0000-0002-6558-1681]{V. Graber} 
\affiliation{Institute of Space Sciences (ICE, CSIC), Campus UAB, Carrer de Can Magrans s/n, E-08193, Barcelona, Spain}
\affiliation{Institut d’Estudis Espacials de Catalunya (IEEC), Carrer Gran Capità 2-4, E-08034 Barcelona, Spain}

\author[0000-0001-5902-3731]{A. Possenti}
\affiliation{INAF–Osservatorio Astronomico di Cagliari, Via della Scienza 5, 09047 Selargius, CA, Italy}

\author[0000-0003-0554-7286]{C. Dehman} 
\affiliation{Institute of Space Sciences (ICE, CSIC), Campus UAB, Carrer de Can Magrans s/n, E-08193, Barcelona, Spain}
\affiliation{Institut d’Estudis Espacials de Catalunya (IEEC), Carrer Gran Capità 2-4, E-08034 Barcelona, Spain}

\author[0000-0003-2781-9107]{M. Ronchi} 
\affiliation{Institute of Space Sciences (ICE, CSIC), Campus UAB, Carrer de Can Magrans s/n, E-08193, Barcelona, Spain}
\affiliation{Institut d’Estudis Espacials de Catalunya (IEEC), Carrer Gran Capità 2-4, E-08034 Barcelona, Spain}

\author[0000-0001-5126-1719]{S. Loru}
\affiliation{INAF -- Osservatorio Astrofisico di Catania, Via Santa Sofia 78, I--95123 Catania, IT}

\begin{abstract}
\noindent
\srclong\ is a radio-loud magnetar with a spin period of 1.36\,s and a dipolar magnetic field strength of $B\sim3\times10^{14}$\,G, which is very young compared to the Galactic pulsar population. 
We report here on the long-term X-ray monitoring campaign of this young magnetar using \xmm, \nustar,\ and \swift\ from the activation of its first outburst in March 2020 until October 2021, as well as INTEGRAL upper limits on its hard X-ray emission. 
The 1--10\,keV magnetar spectrum is well modeled by an absorbed blackbody with a temperature of $kT_{\rm BB}\sim1.1$\,keV, and apparent reduction in the radius of the emitting region from $\sim$0.6 to $\sim$0.2\,km. We also confirm the bright diffuse X-ray emission around the source extending between $\sim$50\arcsec\ and $\sim$110\arcsec. A timing analysis revealed large torque variability, with an average spin-down rate $\dot{\nu}\,\sim\,-$2.3$\,\times\,10^{-11}\,$Hz$^2$ that appears to decrease in magnitude over time. We also observed \srclong\, with the Karl G. Jansky Very Large Array (VLA) on 2021 March 22. We detected the radio counterpart to \src\, measuring a flux density of $S_{v} = 4.38\pm0.05$\,mJy at 3\,GHz, and a half ring-like structure of bright diffuse radio emission located at $\sim$90\arcsec\ to the west of the magnetar. We tentatively suggest that the diffuse X-ray emission is due to a dust scattering halo and that the radio structure may be associated with the supernova remnant of this young pulsar, based on its morphology.
\end{abstract}




\section{Introduction}\label{sec:intro}

Magnetars are a sub-group of isolated neutron stars with ultra-high magnetic fields of $B \approx 10^{14}-10^{15}$\,G. As suggested by \citet{duncan92}, the decay of their extremely strong magnetic fields in the interior of the star is the main energy source of their electromagnetic radiation  \citep[for recent reviews see e.g.][]{kaspi17, esposito21}.  

The observational characteristics of magnetars place them in the top-right corner of the $P-\dot{P}$ diagram (where $P$ is the period and $\dot{P}$ is the period derivative), with relatively long spin periods in the range of 0.3 -- 12\,s and spin-down rates between $\dot{P} \sim 10^{-13}-10^{-11}$\,\ss . Generally, magnetars show bright soft X-ray emission with luminosities in the range of $L_{X} \approx 10^{31} - 10^{36}$\,\lum, sometimes reaching into the hard X-ray energies. The soft X-ray spectrum of magnetars (0.5--10\,keV) typically consists of: i) a thermal component that is usually well modeled by a blackbody with a temperature of $kT_{BB} \sim 0.3 - 1 $ keV; ii) a non-thermal component that can be described by a power-law with a photon index $ \Gamma \sim 2 - 4$. If emitting at hard X-rays, their spectra above 10\,keV are non-thermal, and well modeled with a power-law \citep[][]{turolla15, kaspi17, esposito21}.

\begin{figure}[t]
\resizebox{\hsize}{!}{\includegraphics[angle=0]{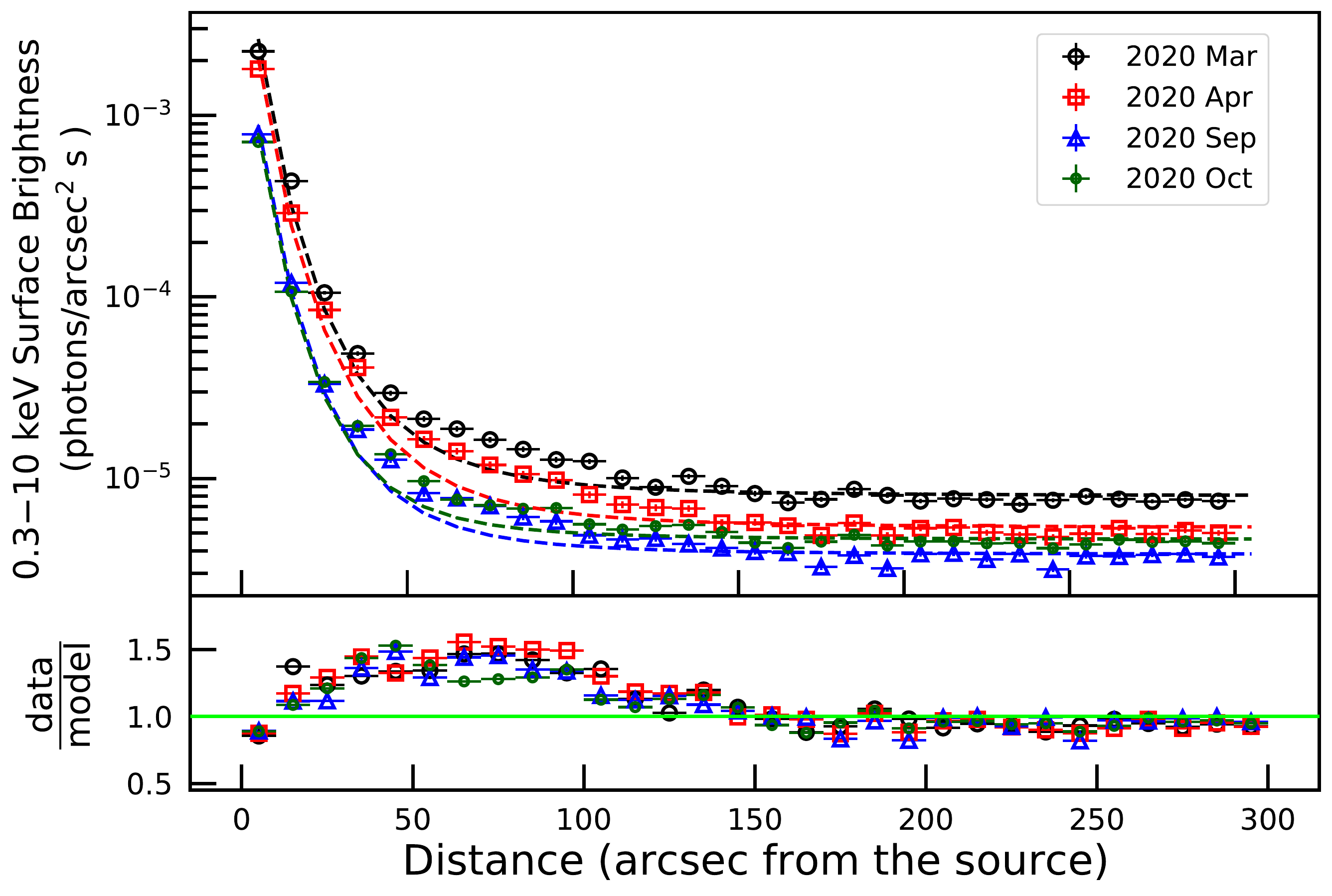}}
\resizebox{\hsize}{!}{\includegraphics[angle=0,trim={0 3.cm 1.8cm 3.2cm},clip]{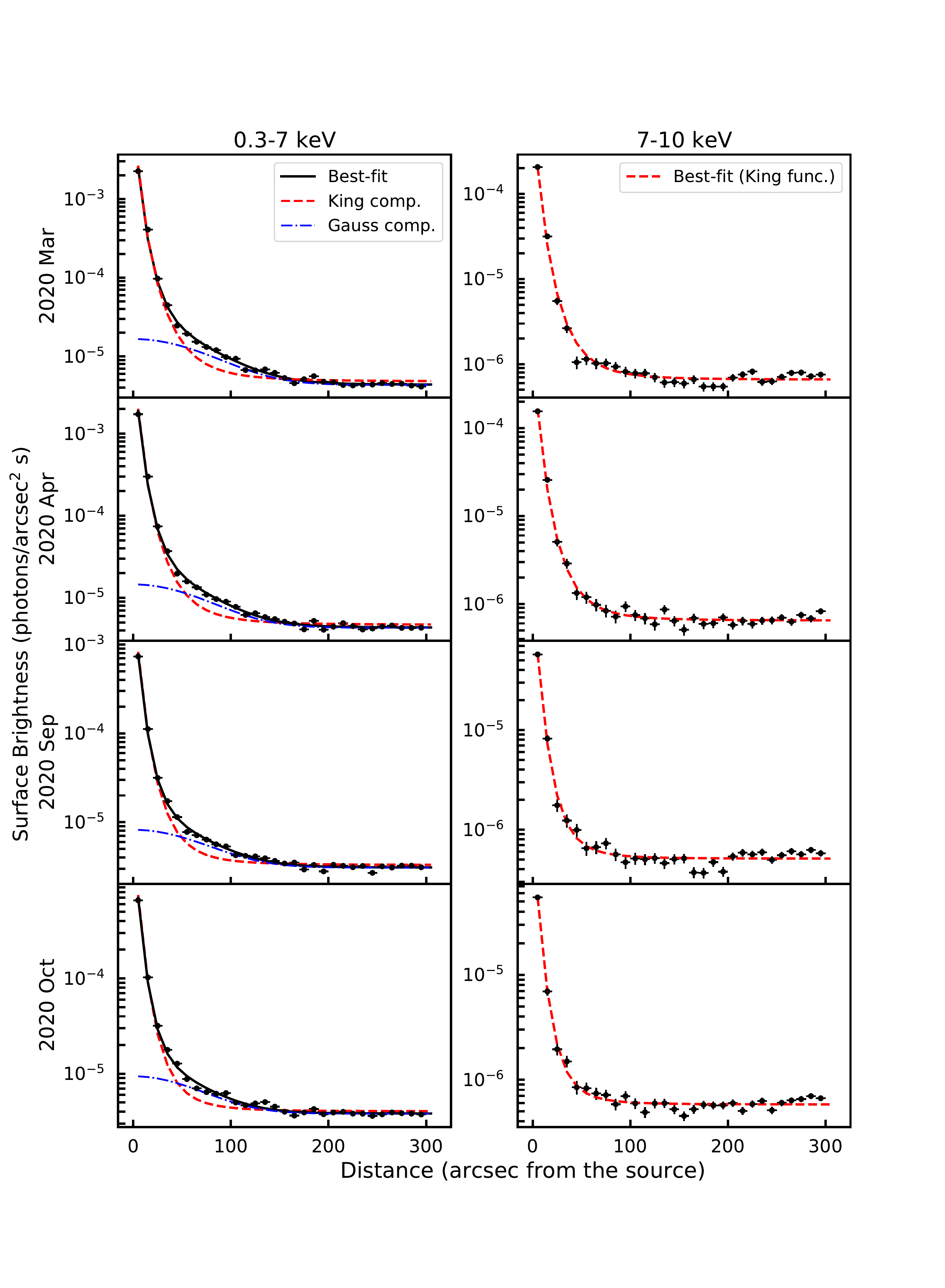}}
\vspace{-0.1cm}
\caption{\label{diff_profile} {\it Top}: Observed X-ray surface brightness up to a radial distance of 300 arcsec in the 0.3--10\,keV energy range extracted from the four \xmm\ observations (the error bars are smaller than the size of the markers). The dashed lines represent the best-fit PSF model. The ratio between the data and the best-fit model is plotted in the bottom panel. {\it Bottom}: Observed X-ray surface brightness up to a radial distance of 300 arcsec in two different energy bands, 0.3--7\,keV (left) and 7--10\,keV (right). The best-fitting models are superimposed.}
\end{figure}

\begin{figure}[t]
\resizebox{\hsize}{!}{\includegraphics[angle=0]{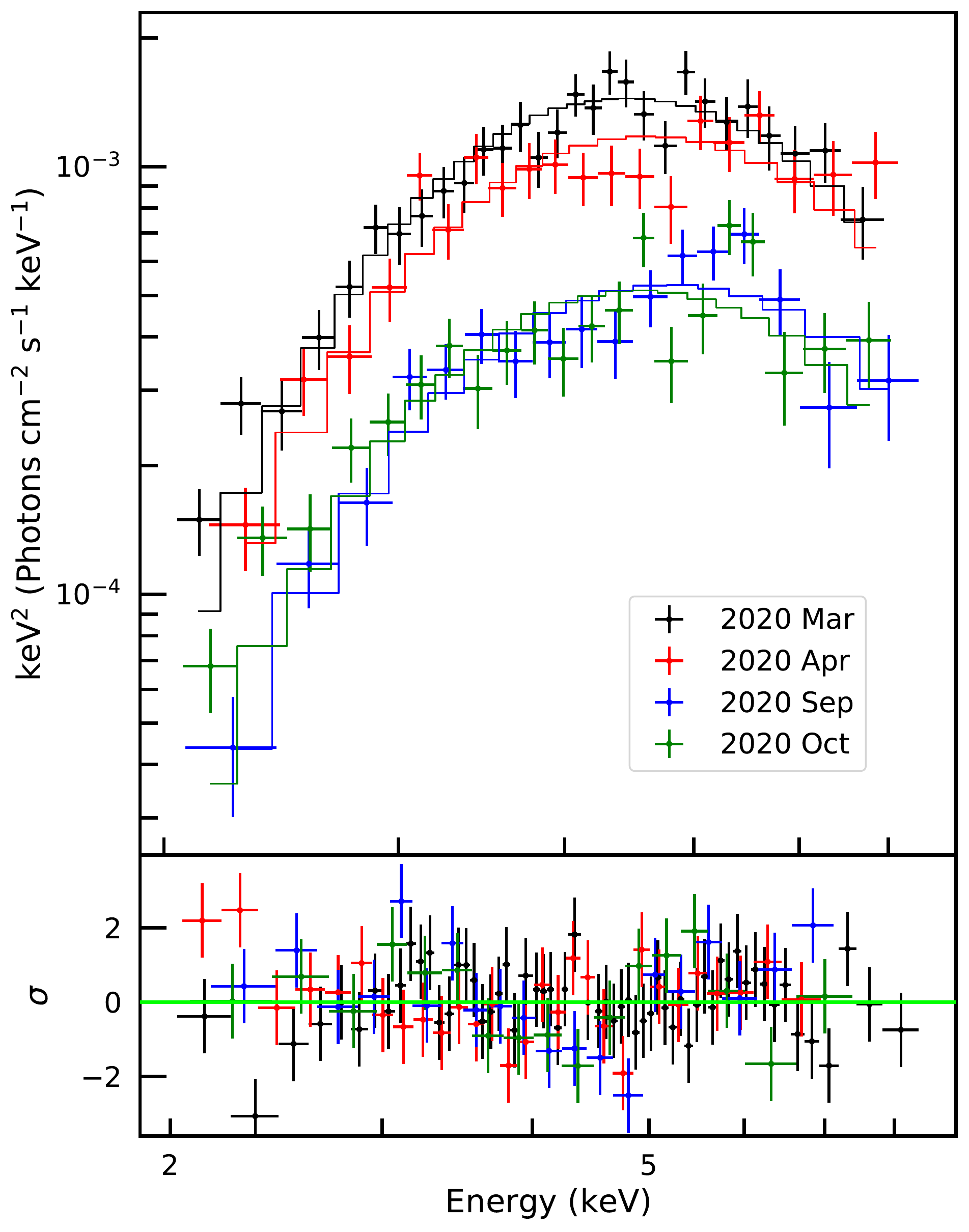}}
\vspace{-0.1cm}
\caption{\label{fig:diffusespec} $E^2 f(E)$ unfolded spectra of the diffuse emission extracted from the four \xmm\ observations. The solid lines mark the absorbed blackbody model. Post-fit residuals in units of standard deviations are shown in the bottom panel.}
\end{figure}

In addition, magnetars exhibit several types of transient activity ranging from short-lived X-ray bursts (time scales of milliseconds to seconds) to longer duration events called Giant Flares (time scale of seconds to tens of minutes). The luminosities of these flares range between 10$^{39}$–10$^{47}$ \lum. Furthermore, magnetars exhibit so-called outbursts, during which their persistent X-ray fluxes suddenly increase by a factor of 10--1000 and then gradually decay over months to years time-scale \citep[][see the Magnetar Outburst Online Catalog\footnote{\url{http://magnetars.ice.csic.es/}}]{Rea2011, cotizelati18}. \\
To date, pulsed radio emission has been detected in 6 magnetars and has been associated with X-ray outbursts in most cases\footnote{We include SGR\,J1935+2154 in the sub-group of radio loud magnetars. For this source, the detection of a periodic radio emission by FAST was claimed in the aftermath of a radio bursting period in 2020 October \citep{zhu20}.}. This emission is characterized by variable radio flux and spectra, and bright single pulses. The magnetar SGR\,J1935+2154 emitted a radio burst with properties similar to those of fast radio bursts (FRBs) during the early stage of its 2020 outburst \citep{chime20, bochenek20}. 
FRBs are bright radio pulses characterised by a millisecond duration and dispersion measures greater than the Galactic one, suggesting an extragalactic origin \citep[for a detailed review see][and references therein]{petroff}. The progenitor engines of FRBs are still broadly discussed in the literature. However, the detection of FRB-like bursts from SGR\,J1935+2154 supports the scenario that magnetars can power at least a subgroup of FRBs.

On 2020 March 12, the Burst Alert Telescope (BAT) on board the {\it Neil Gehrels Swift Observatory} \citep{gehrels04} triggered on a short burst of $\sim$ 0.1 sec, which led to the discovery of a new magnetar \citep{evans20}.
Following this trigger, 64.2 seconds later, the {\it Swift} X-ray Telescope (XRT) started observing the field and reported a new X-ray source, \srclong{} (hereafter, \src).
Four hours after the \swift/BAT alert, the Neutron star Interior Composition Explorer ({\it NICER}) started a series of observations of the source which revealed a coherent periodicity of 0.733417(4)~Hz \citep{enoto20}. The {\it NICER} periodicity and the magnetar-like burst detected by \swift/BAT suggested that \src{} is a new fast-spinning magnetar with a spin period of 1.36 s.

Follow-up radio observations performed by the 100-m Effelsberg radio telescope and the 76-m Lovell Telescope detected radio pulsations at a frequency of 0.7334110(2)~Hz, confirming \src{} as a radio-loud magnetar \citep{karuppusamy20}. The radio monitoring campaign provided a measurement of the spin-period derivative of $\dot{P} = 8.16(2) \times 10^{-11}$\,\ss , resulting in a first estimate of the dipolar surface magnetic field at the equator of $B \sim 3.4 \times 10^{14}$\,G and a characteristic age of $\sim$ 265 yr \citep{karuppusamy20, ccc+20, esposito20}. Even from these early estimates of the timing parameters, it was clear that this new magnetar is very young compared to the rest of the magnetar population. 
Additionally, the dispersion measure DM = 706(4)\,cm$^{-3}$\,pc suggested a source distance of 4.8 or 8.1\,kpc, depending on the model used for the Galactic free electron density  \citep{karuppusamy20, ccc+20}.

Since its discovery, several X-ray \citep{esposito20,hu20,blumer20} and radio telescopes \citep{karuppusamy20, champion20,lower20a,lower20b,huang21, rajwade22} have monitored this young magnetar during the evolution of its outburst, confirming its noisy spin-period evolution and X-ray outburst decay. Here we report on follow-up observations with \xmm, \nustar, \swift\ and INTEGRAL to study the X-ray spectral and timing evolution of \src\, along the decay of its first outburst, covering $\sim$19 months since the outburst onset. Furthermore, we report on radio continuum observations performed with the VLA, that allowed us to search for the supernova remnant around this young pulsar, leftover of the ejected materials after the supernova explosion \citep[see full review][]{vink12}. 

We describe the observations and data reduction in Section \ref{sec:data}. In Section \ref{sec:analysis}, we introduce the X-ray spectral analysis for the diffuse emission observed around the magnetar (Sec.\,\ref{subsec:diffuse}) and for the magnetar itself (Sec.\,\ref{subsec:specatral}), a burst search (Sec.\,\ref{subsec:burst}), the X-ray timing analysis (Sec. \ref{subsec:timing}) and the analysis of radio continuum data (Sec.\,\ref{subsec:radio_emission}). Finally, we discuss the results in Section \,\ref{sec:disc}.


\section{X-ray Observations and data reduction}
We report the log of the observations used in this work in Table \ref{tab:observations1}.
We performed the data reduction using the \textsc{heasoft}\footnote{ \url{https://heasarc.gsfc.nasa.gov/docs/software/heasoft/}} package \citep[v.6.29c;][]{heasoft14}.
All uncertainties in the text are reported at 1$\sigma$ confidence level, unless otherwise specified. Throughout this work, we adopt a distance of 4.8\,kpc \citep{karuppusamy20}.

\label{sec:data}
\subsection{XMM-Newton}
\src\ was monitored four times with the European Photon Imaging Camera (EPIC) on board the \xmm\ satellite between 2020 March 15 and October 8 for a total exposure time of $\sim$137\,ks. The exposures ranged from 22\,ks to 49\,ks (Table\,\ref{tab:observations1}). The EPIC-pn \citep{struder01} was set in large window mode (LW; timing resolution of 47.7\,ms) for the first observation and in full frame mode (FF, timing resolution of 73.4\,ms) for the remaining observations, while both Metal Oxide Semi-conductor cameras (MOS) \citep{turner01} were operating in small window mode (SW; timing resolution of 0.3\,s). 
Raw data were analyzed with the \textsc{sas}\footnote{\url{ttps://www.cosmos.esa.int/web/xmm-newton/sas}} software package \citep[v.19.1.0;][]{gabriel04}. 
We cleaned the observations from periods of high background activity. This resulted in a net exposure time of 11.2, 9.5, 16.4 and  19.6 ks for the four observations ordered chronologically.
No pileup was detected. We selected the source photon counts from a circle of 30\arcsec\ radius, while the background level was estimated from a circle of 100\arcsec\, radius, on the same CCD away from the source.
For the diffuse emission, we extracted the spectrum by selecting source photon counts from an annulus of radii 50\arcsec--110\arcsec, centered on the source, and used the same background region as adopted for the point-like source (more details in subsection\,\ref{subsec:diffuse}). We focused this study on the EPIC-pn data, but checked that MOS data gave consistent results.

\subsection{NuSTAR}
The Nuclear Spectroscopic Telescope ARray \citep[{\it NuSTAR};][]{harrison13} observed \src\ six times starting on 2020 March 3 and ending on 2020 September 7 for a total exposure time of $\sim$180\,ks (see Table~\ref{tab:observations1}). The longest exposure was 59\,ks taken under Obs.ID 80402308004, while the shortest exposure was 12\,ks for Obs.ID 80402308010. 
The source photon counts were extracted from a circle of radius 100 arcsec in the first three observations and from a smaller circle in the following three observations (the adopted radii varied in the range 50--80 arcsec, mainly depending on the presence of significant stray light contamination near the source position). The background level was estimated from a circle of radius 100 arcsec located near the source position in all cases. 

\subsection{Swift}
\swift/XRT \citep{burrows05} extensively monitored \src\, since the outburst onset until the end of October 2021. The observations were carried out either in photon counting mode (PC, time resolution of 2.51\,s) or in windowed timing mode (WT, time resolution of 1.779\,ms). Data were reprocessed using standard prescriptions and software packages such as \textsc{xrtpipline\footnote{\url{https://www.swift.ac.uk/analysis/xrt/xrtpipeline.php}}}. Source counts were accumulated within a circular region of radius of 20 pixels (1 XRT pixel corresponds to 2\farcs36), while the background photons were extracted from an annulus with radii of 100--150 pixels and from a 20-pixel-radius circle for PC-mode and WT-mode observations, respectively.

\subsection{INTEGRAL}
\INT\, \citep{winkler03} observed \src\, between 2020 March 13 at 21:22:56 UT and 2020 March 16 at 03:47:32 UT as part of our approved magnetar ToO program, for a total exposure time of about 105\,ks. 
Unfortunately, its soft X-ray spectrum did not allow detection in the hard X-ray band. We derived 3$\sigma$ upper limits on the observed flux with ISGRI \citep{lebrun2003} at the level of 1.8 $\times$ 10$^{-11}$\,\flux\ (28--40\,keV) and 3.5 $\times$ 10$^{-11}$\,\flux\ (40--80\,keV).

\section{Analysis and results}
\label{sec:analysis}


\subsection{Diffuse Emission}
\label{subsec:diffuse}

The diffuse emission around \src\ has been investigated in a number of previous studies. Using an \xmm\, observation performed a few days after the outburst onset, \cite{esposito20} reported the detection of diffuse emission extending between 50\arcsec\,-- 110\arcsec\ around \src. Another study by \cite{blumer20} showed the presence of diffuse emission also on smaller angular scales, up to 10\arcsec\ using \cxo\ observations. 
 
To constrain the spatial extent of the large-scale diffuse emission, we extracted the radial profile of the observed surface brightness up to a distance of 300 arcsec away from the source for all four \xmm\ observations. We fit it using a King function reproducing the EPIC-pn point-spread function (PSF; \citealt{ghizzardi02}) with the addition of a constant term to model the background level. 
We found a photon excess associated with the diffuse emission at radial distances within $\sim 50-110$ arcsec in all four pointings (see Figure\,\ref{diff_profile}, top panel).\\
To further investigate the energy dependency of the diffuse structure, we built surface brightness profiles in two different energy bands, 0.3--7\,keV and 7--10\,keV (see Figure\,\ref{diff_profile}, bottom panel). 
In the soft energy interval, we included a Gaussian function in the above-mentioned model in order to properly describe the observed photon excess, while this component is not required in the hard band ($F$-test probability $>$0.001 for its inclusion).

We extracted the 0.2--7.5\,keV spectra associated with the diffuse emission by selecting photons within an annulus centered on \src\ with radii of 50\arcsec\ and 110\arcsec\, respectively, and grouped them using the \textsc{specgroup} tool to have a minimum bin size of 100 counts per bin. The ancillary response files for the diffuse emission spectra were generated using the \textsc{arfgen} tool with the \texttt{extendedsource} parameter set to \texttt{yes}, while the redistribution matrix files were created via the \textsc{rmfgen} script. 

To study the spectral behavior of the diffuse component, we performed a simultaneous fit of the source and diffuse emission spectra obtained from the four EPIC-pn observations. The former were described by an absorbed two-blackbody model, while for the latter we adopted a single absorbed blackbody. The hydrogen column density \nh\ was quantified with the \textsc{tbabs} model with the abundance of the interstellar medium taken from \cite{wilms00} and the photoionization
cross-section model from \cite{verner96}. In the fits, the \nh\ was forced to be the same among all the data sets.
At each epoch, the diffuse emission temperature and normalization were tied up between the source and diffuse emission spectra.
Overall, the fit gave a satisfactory description of the data with \nh\ = (1.23$\pm$0.02)$\times$10$^{23}$\,\cm\ and a reduced chi-squared \rchisq=1.2 for 332 degrees of freedom (d.o.f.). The best-fitting values for the temperature $kT_{\rm diff}$ and the observed 0.3--10\,keV flux $F_{\rm X.diff}$ for the diffusion emission are listed in Table\,\ref{tab:spec_parameters}. Figure\,\ref{fig:diffusespec} shows the unfolded spectra of the diffuse emission at the four different epochs with the best-fitting model, marked by a solid line, and the residuals with respect to this model.
 
We detected a flux reduction of the diffuse X-ray emission of about 35\% (decreasing from 1.9 to 0.6 $\times 10^{-11}$\,\flux)  between 2020 March and October. The large flux variability and the soft X-ray spectrum suggest a dust scattering-halo as the source of this diffuse emission.

\begin{figure}[t]
\includegraphics[angle=0, width=0.45\textwidth]{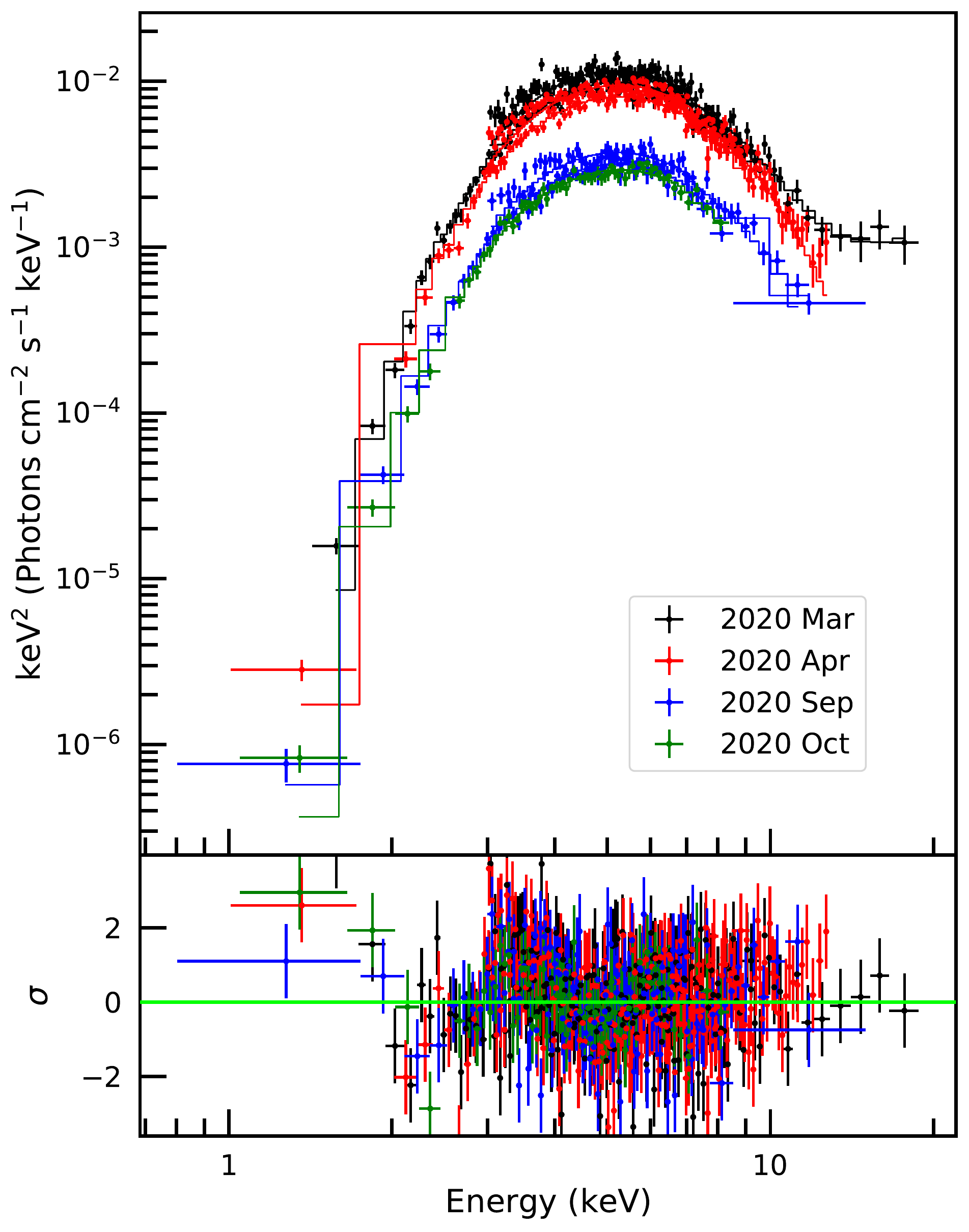}
\caption{\label{fig:spectral} $E^2 f(E)$ unfolded spectra of \src\ from \xmm\ and \nustar\ observations. In the fit we adopted a model consisting of two blackbodies, except for the 2020 March spectra where we included a power law (\rchisq{} = 1.41 for 607 d.o.f). Post-fit residuals in units of standard deviations are shown in the bottom panel.\label{fig:general}} 
\end{figure}


\begin{deluxetable*}{cccccccc}[htb]
\tablecaption{Results of the joint fit of the \xmm\ and \nustar\ spectra of \srclong.
\label{tab:spec_parameters}}
\tabletypesize{\scriptsize}
\tablecolumns{8}
\tablewidth{0pt}
\tablehead{
\colhead{Instrument} & \colhead{Obs.ID} & 
\colhead{$F_{\rm X.obs}$\tablenotemark{{\scriptsize a}} } &
\colhead{$F_{\rm X.unabs}$\tablenotemark{{\scriptsize a}} } &
\colhead{$kT_{\rm BB}$} & 
\colhead{$R_{\rm BB}$ \tablenotemark{{\scriptsize b}} } &
\colhead{$kT_{\rm diff}$\tablenotemark{{\scriptsize c}}} & 
\colhead{$F_{\rm X.diff}$\tablenotemark{{\scriptsize a}}} \\  
\colhead{} &  \colhead{} &
\multicolumn2c{($\times 10^{-11}$\,\flux)} &
\colhead{(keV)} & 
\colhead{(km)} &
\colhead{(keV)} &
\colhead{($\times 10^{-12}$\,\flux)}
}
\startdata
\emph{XMM}/EPIC-pn & 0823591801$^{d1}$ & 1.414 $\pm$ 0.003 & 3.036 $\pm$ 0.005  & 1.13 $\pm$ 0.01    & 0.56 $\pm$ 0.01 & 0.86 $\pm$ 0.02  & 1.93 $\pm$ 0.01\\
\nustar/FPMA & 80402308002$^{d1}$ &  -  & -   & - &- &- & -\\
\emph{XMM}/EPIC-pn & 0823593901$^{d2}$& 1.142 $\pm$ 0.003 & 2.417 $\pm$ 0.005  & 1.149 $\pm$ 0.007  & 0.501 $\pm$ 0.009 & 0.91 $\pm$ 0.03& 1.59 $\pm$ 0.02\\ 
\nustar/FPMA & 80402308004$^{d2}$ &  -  & - &  &- &-  &- \\
\nustar/FPMA & 80402308006  & 1.098 $\pm$ 0.005 & 2.264 $\pm$ 0.007  & 1.19  $\pm$ 0.01 & 0.44 $\pm$ 0.01 & - &-  \\
\nustar/FPMA & 80402308008  & 0.652 $\pm$ 0.008 & 1.43 $\pm$ 0.01  & 1.14  $\pm$ 0.02   & 0.35 $\pm$ 0.02 & - &-  \\
\nustar/FPMA & 80402308010  & 0.70 $\pm$ 0.01   & 1.42 $\pm$ 0.01  & 1.19  $\pm$ 0.03   & 0.36 $\pm$ 0.02 & - &-  \\
\emph{XMM}/EPIC-pn & 0823594001$^{d3}$& 0.448 $\pm$ 0.004  & 0.951 $\pm$ 0.007  & 1.13  $\pm$ 0.01 & 0.320 $\pm$ 0.008& 0.96 $\pm$ 0.04 & 0.72 $\pm$ 0.02\\
\nustar/FPMA & 80402308012$^{d3}$ &  - & -  & -   & -  &-  &- \\
\emph{XMM}/EPIC-pn & 0823594201  & 0.416 $\pm$ 0.006  & 0.876 $\pm$ 0.007  & 1.17 $\pm$ 0.02 & 0.28 $\pm$ 0.01 & 0.89 $\pm$ 0.03  & 0.68 $\pm$ 0.02\\ 
\hline
\hline
\enddata
\tablenotetext{a}{The fluxes are measured in the 0.3--10\,keV energy range.}
\tablenotetext{b}{The blackbody radius is derived assuming a source distance of 4.8\,kpc.}
\tablenotetext{c}{The fit of the diffuse emission is performed in the 2--7.5\,keV energy range.}
\tablenotetext{d}{These observations were fitted simultaneously.}
\end{deluxetable*}

\begin{figure*}
\resizebox{\hsize}{!}{\includegraphics[angle=0]{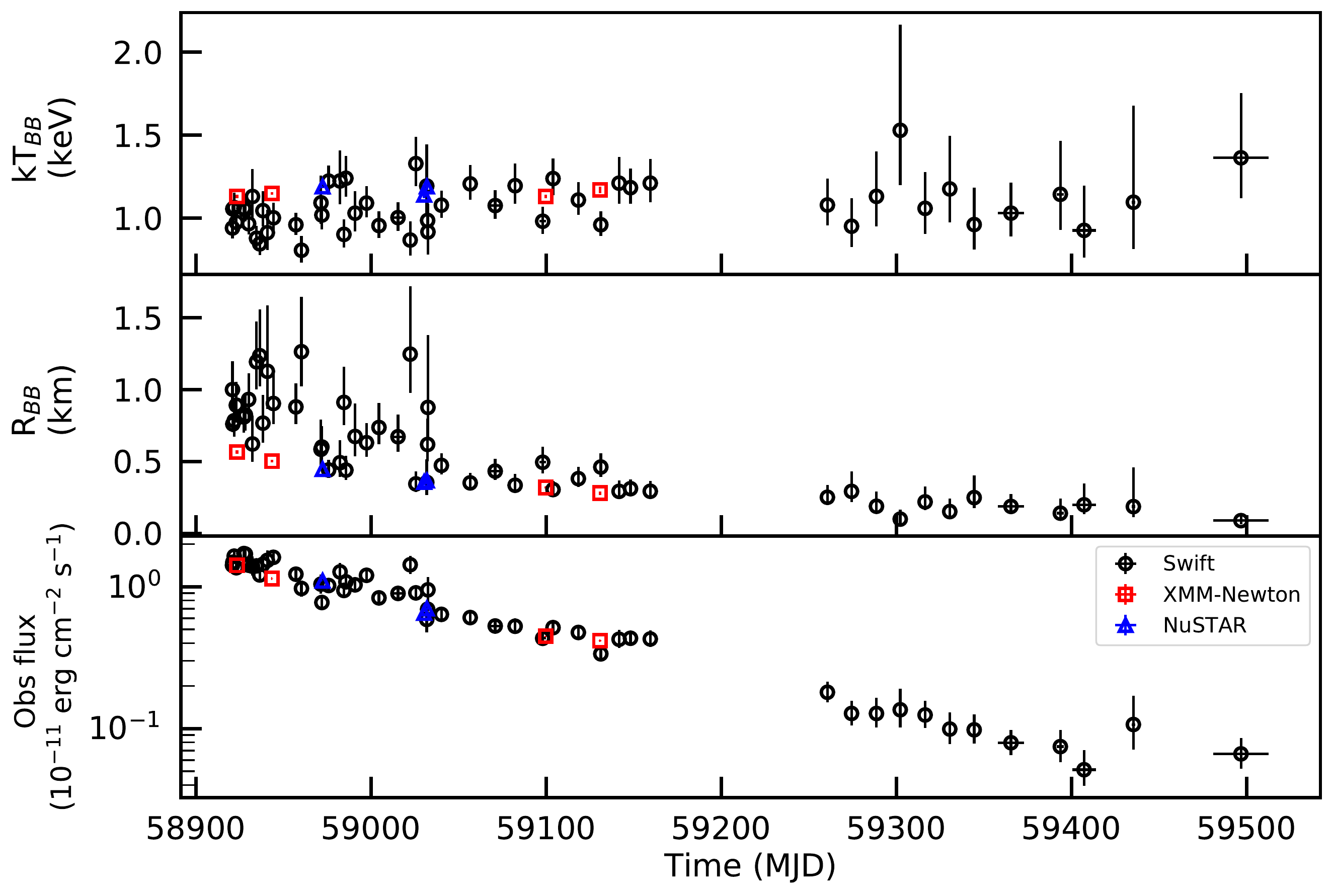}}
\caption{\label{lcurve} Temporal evolution of the blackbody temperature (top) and radius (middle). The latter is evaluated for a distance of 4.8\,kpc (see Sec.\,\ref{subsec:specatral} for more details). The bottom panel shows the temporal evolution of the observed flux in the 0.3--10\,keV energy range.}
\end{figure*}

\subsection{Spectral Analysis}
\label{subsec:specatral}

We used the \xmm\ and \nustar\ data to study the X-ray emission of \src\ from soft to hard X-rays.
For the \xmm/EPIC-pn observations, we grouped the X-ray spectra using the \textsc{specgroup} tool to have a minimum bin size of 100 counts per bin. For the \nustar/FPMA observations, the spectra were grouped with \textsc{grppha} to have a minimum bin size of 50 counts per bin.

The spectral fitting was performed using \textsc{Xspec} \citep[v12.11.1;][]{arnaud96}. To model the spectra, we selected the 3--13\,keV energy range for all \nustar\ spectra, except for the first spectrum (Obs.ID 80402308002) which was modeled in the 3--20\,keV energy range. We restricted the energy range of the EPIC-pn spectra  to 1--10\,keV due to the domination of the background below 1\,keV for this highly absorbed source. To quantify the hydrogen column density \nh, we adopted the \textsc{tbabs} model with the interstellar-radiation abundance from \cite{wilms00}, and the photoionization cross-section model from \cite{verner96}.

We modeled the EPIC-pn and FPMA spectra of \src\ simultaneously with two blackbodies plus a power-law component. We fixed the temperature and normalization of the first blackbody component to the aforementioned values derived from the diffuse emission fit (see Section \ref{subsec:diffuse} and Table\,\ref{tab:spec_parameters}). 
We also added a constant term between the two instruments to account for cross-calibration uncertainties. The constant was fixed to one for the \xmm/EPIC-pn spectra and left free to vary for the \nustar/FPMA spectra. We linked the hydrogen column density between all the spectra and obtained \nh $= (1.24 \pm 0.02)\times10^{23}$\,cm$^{-2}$. We found that the model fits the data with \rchisq=1.41 for 607 d.o.f. (see Figure \ref{fig:spectral}). We note that the power-law component is required only for the spectrum at the outburst peak, i.e, the 2020 March 15 epoch, thus we did not include this component in the model for the remaining spectra. We obtained a power-law photon index of $\Gamma = 1.0 \pm 0.6$, 
which is consistent with the result reported by \cite{esposito20}, $\Gamma = 0.0 \pm 1.3$. In Table\,\ref{tab:spec_parameters}, we list the best-fit parameters for the blackbody temperatures $kT_{\rm BB}$ and radii $R_{\rm BB}$, as well as the observed $F_{\rm X.obs}$ and unabsorbed $F_{\rm X.unabs}$ fluxes estimated in the 0.3--10\,keV energy interval.

To supplement the \xmm\ and \nustar\ observations, we initiated a \swift/XRT monitoring campaign of \src. These pointings were used to sample the flux and spectral evolution of the magnetar over a longer time-span. We fit all the \swift\ spectra simultaneously with an absorbed blackbody model fixing \nh\ to the value obtained from the broad-band fit with \xmm\ and \nustar\ data. Figure\,\ref{lcurve} shows the temporal evolution of the blackbody temperature and radius, and the 0.3--10\,keV observed flux from the outburst onset on 2020 March 12 until 2021 October 24. The 0.3--10\,keV observed flux of \src\ has shown a rapid decay since the outburst peak, from $\sim$1.4 $\times$ 10$^{-11}$\,\flux\ to $\sim$6.6 $\times$ 10$^{-13}$\,\flux\ after about 19 months (see Figure\,\ref{lcurve}, bottom panel). Similarly, the blackbody radius decreased from $\sim$0.6\,km to $\sim$0.3\,km during the first seven months and then settled at an average value of $\sim$0.2\,km (middle panel). We did not observe significant variability in the blackbody temperature, which attained a constant value of $\sim$1.1\,keV over the whole monitoring campaign (top panel).

\begin{figure}[]
    \includegraphics[scale=0.53]{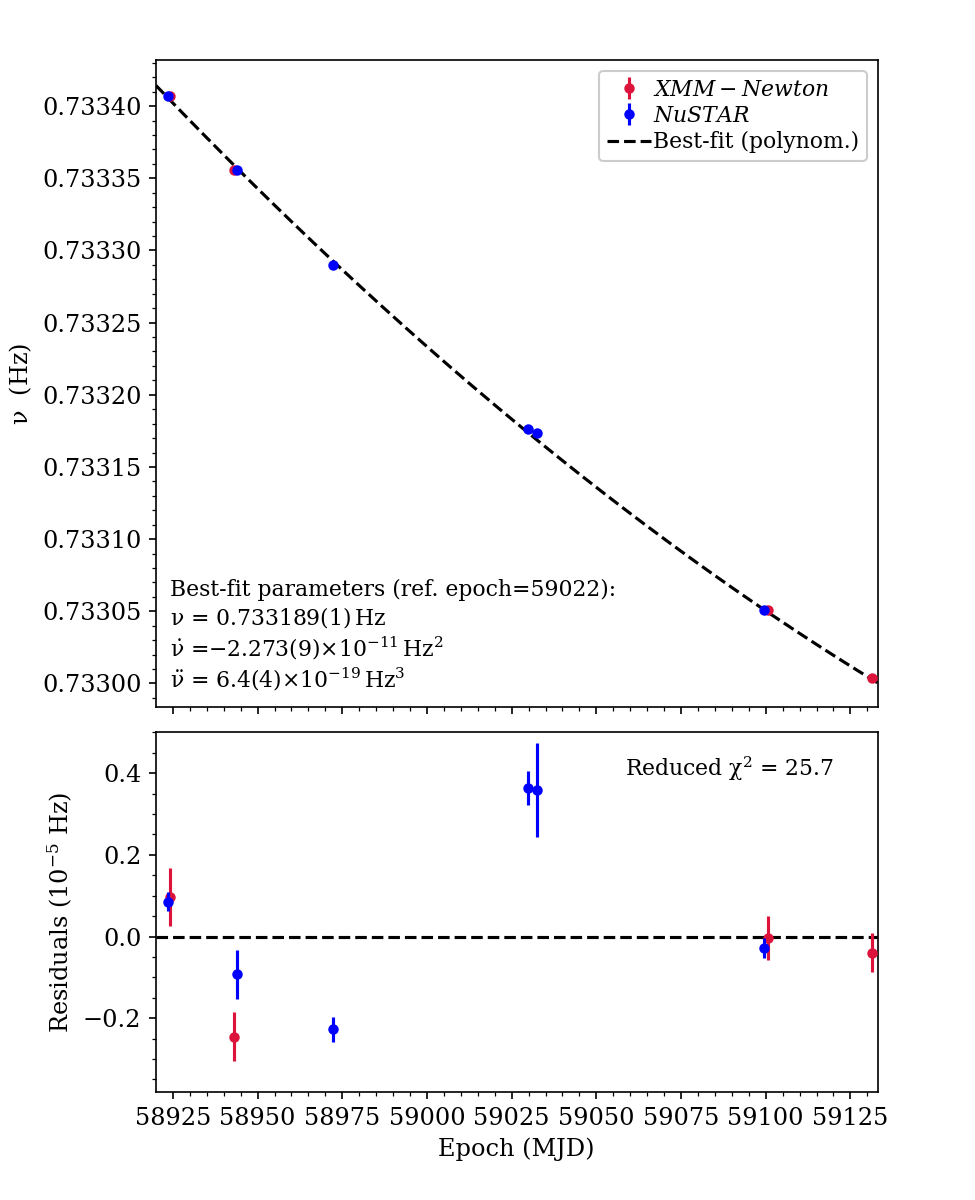}
    \caption{\texttt{TEMPO}-derived values of the spin frequency, $\nu$, in each observations. The epochs of \xmm\ and \nustar\ observations are shown as red and blue points. Error bars correspond to the 1-$\sigma$ uncertainties reported by \texttt{TEMPO}. Black dashed/dotted lines are the best-fit models for $\nu$(t) (see text in section \ref{subsec:timing}). As evidenced by the large reduced $\chi^2$ values we measured (provided in the top-right corner of the bottom panels), the $\nu$(t) solutions poorly fit the data. This is explained by the large timing noise present in \src\ and the simplicity of our models.}
    \label{fig:timing2}
\end{figure}

\subsection{Burst Search \label{subsec:burst}}

The sky region of \srclong\, has been extensively observed every year by the \INT\, satellite, starting from March 2003.
We have carried out a search for bursts from \src\ using the data of IBIS/ISGRI, a coded mask imaging instrument with  angular resolution of $\sim$12 arcmin and  field of view of  $29^{\circ}\times29^{\circ}$ \citep{ubertini2003}. We selected all the public data with \src\ in the field of view obtained until April 2021. After removal of time intervals with high and variable background, this amounted to an exposure time of about 43\,Ms. Most of the considered data (41\,Ms) were obtained before the discovery outburst. 

The burst search was done with the procedure described by \cite{mereghetti2021}. Briefly, this consists in a first screening of the light curves binned at eight logarithmically  spaced  timescales from 0.01 s to 1.28 s to select excesses with respect to the locally measured background count rate. The search is carried out in the nominal 20-100 keV, 30-100 keV and 30-200 keV energy ranges. In this first step, a threshold corresponding to $\sim$0.001 false positives per Science Window and time scale is adopted (\INT\, data are divided in Science Windows of a few ks duration).  All these excesses are  then examined through an imaging analysis, in order to reject the events caused by instrumental background or by bright sources located outside the field of view. This procedure led to the (re)-discovery of several bursts from other sources in the field of view, most of which originating from the magnetar SGR 1806-20 \citep{gotz2006}, located at about 5 degrees from \src. However, no significant bursts were found from \src, with a 3$\sigma$ upper limit on the 20--200\,keV fluence of about $10^{-8}$\,erg\,cm$^{-2}$. 

We also performed a burst search on \xmm\ and \nustar\ data, using the method described by \citet{borghese20}; (see also, e.g., \citealt{gkw04}). We built the source barycentered light curves with time resolutions of 1/16, 1/32 and 1/64\,s. We tagged as bursts the bins with a probability $<$10$^{-4}$($NN_{\rm trials}$)$^{-1}$, where $N$ is the total number of time bins in a given light curve and $N_{{\rm trials}}$ corresponds to the number of timing resolutions used in the search. No bursts were detected in the \xmm/EPIC-pn light curves, while we list the epochs of the bursts found in the \nustar\ datasets in Table\,\ref{tab:bursts}. Due to the low photon statistics, we were unable to model the corresponding spectra.



\subsection{Timing Analysis\label{subsec:timing}}
For the timing analysis of \src, we used the \xmm\ and \nustar\ data sets as shown in Table \ref{tab:observations1}. For \nustar, we applied the clock corrections with up-to-date clock files and combined the FPMA and FPMB events files for each observation.
For both \xmm\ and \nustar\ data sets, we referred the photon arrival times to the Solar system barycentre adopting the source coordinates by \cite{esposito20}, i.e.,  R.A.= 18$^h$ 18$^m$ 00$^s$.16, Dec.= --16\textdegree $07' 53.2''$ (J2000.0), and the JPL planetary ephemeris DE200.

\begin{deluxetable}{rcr}
\tablecaption{ Best-fit spin frequencies  calculated with \texttt{TEMPO} in individual \xmm\ and \nustar\ observations. Numbers in parentheses are the 1$\sigma$ uncertainties on the last digit reported by \texttt{TEMPO}. 
\label{tab:timingepoch}}
\tablecolumns{3}
\tablehead{
\colhead{Instrument/Obs.ID} &
\colhead{Ref. Epoch } &
\colhead{$\nu$} \\[-0.5em]
\colhead{} & 
\colhead{(MJD)} &
\colhead{(Hz)} 
}
\startdata
     {\it XMM/}0823591801 &  58923.40 &  0.7334073(7) \\
 {\it NuSTAR/}80402308002 &	 58923.40 &	 0.7334068(2) \\
     {\it XMM/}0823593901 &	 58943.30 &	 0.733356(6) \\
 {\it NuSTAR/}80402308004 &	 58944.00 &	 0.7333558(6) \\
 {\it NuSTAR/}80402308006 &	 58972.40 &	 0.73329(3) \\
 {\it NuSTAR/}80402308008 &	 59030.40 &	 0.7331763(4) \\
 {\it NuSTAR/}80402308010 &	 59031.90 &	 0.733173(1) \\
 {\it NuSTAR/}80402308012 &	 59099.50 &	 0.7330509(3) \\
     {\it XMM/}0823594001 &	 59099.80 &	 0.7330506(5) \\
     {\it XMM/}0823594201 &	 59130.60 &	 0.7330035(5) \\
\hline
\hline
\enddata
\end{deluxetable}

In an attempt to derive a phase-coherent timing solution, we first used a provisional ephemeris to assign the rotational phase of each photon in the data sets through an unbinned maximum likelihood method \citep{lrc+09,rkp+11}. This was done with the \textsc{photonphase} tool of the PINT\footnote{ \url{https://github.com/nanograv/PINT}} pulsar timing package \citep{2021ApJ...911...45L}. To account for the different energy bandpass, we created two separate template profiles for the \xmm\ and \nustar\ observations by folding the strongest detection in their respective dataset. Two-component Gaussian models were fitted to the binned profiles to construct smoothed standard templates. Times of arrival (TOAs) and associated errors were then computed for a number of sub-integrations from the predicted photon-phase information, using the smoothed profile templates to define the fiducial point in pulse phase.

Using the \textsc{tempo} timing software \citep{2015ascl.soft09002N}, we then tried to obtain a coherent timing solution that simultaneously fits the \xmm\ and \nustar\ TOAs. Similar to the post-outburst spin evolution seen in other magnetars (see review by \citealt{kaspi17}), we also observed significant variability in the spin-down behaviour of \src, particularly a large jump in spin frequency between MJDs 58972 and 59030. This could be an evidence for the presence of discrete timing events and/or strong timing noise, which is consistent with the erratic timing behavior reported by \cite{ccc+20}, \cite{hu20} and \cite{rajwade22}. To account for the large spin variability of \src\, we included up to four spin frequency derivatives in our timing model. However, due to the sparsity of our observations, we encountered phase-count ambiguity during the phase-connection procedure which could not be resolved even with the aid of automated algorithms such as Dracula\footnote{\url{https://github.com/pfreire163/Dracula}} \citep{fr18}. 

Nevertheless, we examined the rotational evolution of \src\ by measuring the spin frequency $\nu$ of the magnetar in each observation using the computed TOAs and \textsc{tempo}. The resulting $\nu$ values are listed in Table~\ref{tab:timingepoch}, and shown in  Figure~\ref{fig:timing2}. We modeled long-term average spin evolution $\nu$(t) with a second-order polynomial function (dashed line in Figure~\ref{fig:timing2}) and the resulting best-fit spin-down rate on MJD 59022 is $\dot{\nu}\,=\,-$2.273(9)$\,\times\,10^{-11}\,$Hz$^2$. However, this simple model poorly fits the data (\rchisq\ of 25.7), and because of the large time gaps between our observations, we cannot determine whether the large residuals are caused by an unmodeled timing anomaly (glitch) or timing noise. 

We also note that we attempted to fold the \xmm\ and \nustar\ data using the timing ephemerides provided by \cite{ccc+20}, \cite{hu20} and \cite{rajwade22}. These ephemerides fail to predict the rotational phase of the source during our \xmm\ and \nustar\ observations, with one exception: the solution from \cite{rajwade22} extrapolates well the rotation of \src\ during our last \xmm\ observation in 2020 September (MJD 59130). This is not surprising considering that large variations in spin down are reported by \cite{ccc+20}, \cite{hu20} and \cite{rajwade22} for MJDs before $\sim$59100, after which the spin down appears to stabilize around a mean value of $\dot{\nu}\,\sim\,-1.37\,\times\,10^{-11}\,$Hz$^{2}$ \citep{rajwade22}.

Considering our limited and sparse dataset, as well as the poorly-understood impact of magnetospheric processes on spin behaviour associated with magnetar outbursts, we do not attempt to further model, quantify and/or interpret the timing properties of \src\ in the \xmm\ and \nustar\ data.

\begin{figure*}[t]
\centering
\includegraphics[width=1.\textwidth]{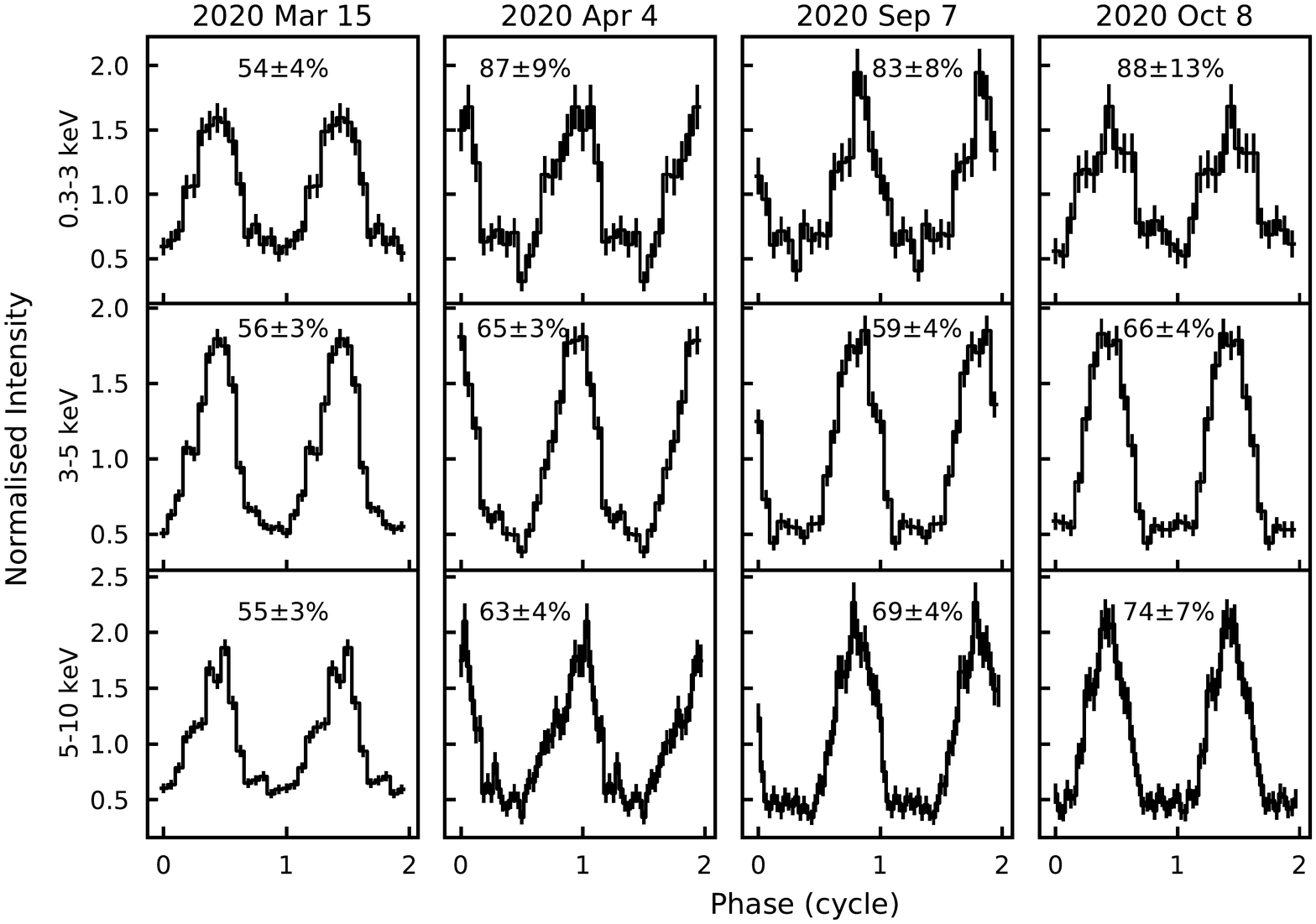}
\caption{Energy-resolved pulse profiles of \src\ extracted from the four \xmm\ data sets presented in this work. The profiles were obtained by folding the light curves using the frequencies reported in Table \ref{tab:timingepoch}. The corresponding pulsed fraction values are reported in each panel. Two cycles are shown for clarity. }
\label{fig:pulse_profile}
\end{figure*} 

To study possible changes in the shape and amplitude of the X-ray pulse profile with photon energy, we extracted energy-resolved pulse profiles from the EPIC-pn data sets in three energy bands: 0.3--3, 3--5 and 5--10\,keV (see Figure \ref{fig:pulse_profile}). 
This is performed by folding the time series on the spin periods reported in Table \ref{tab:timingepoch}. The rows in Figure \ref{fig:pulse_profile} show the evolution of the pulse profile in time, while the columns show their evolution in energy. For each panel, we also reported the corresponding values of the pulsed fraction (PF) that is defined as:

\begin{equation}
PF = \frac{(CR_{\max} - CR_{\min})}{( CR_{\max} + CR_{\min})},
\label{eq:pulsedfraction}
\end{equation}

\noindent
where CR$_{\max}$ and CR$_{\min}$ are the count rates at the maximum and minimum of the pulse profile. For a given energy band the PF increased in time from March to October 2020 epochs. Additionally, we also estimated the PF for the 0.3--10\,keV energy interval: it increased with time, from (53$\pm$2)\% to (64$\pm$3)\% between March and October 2020.


\subsection{Radio observations}
\label{subsec:radio_emission}
We observed \src\ with the VLA under project code 21A-111, with the aim to detect the radio counterpart of \src, as well as the presence of any diffuse radio emission around the source. The VLA observation was performed on 2021 March 22 (MJD~59295) with the telescope on source between 14:58 and 15:38 UT. The data were carried out in the $S$-band, at the central frequency of 3\,GHz and a total bandwidth of 2\,GHz (comprised of sixteen 128-MHz sub-bands made up of sixty-four 2-MHz channels). 3C~286 was used for bandpass and flux calibration, while the nearby (6.5$^{\circ}$ away) source J1822$-$0938 was used for phase calibration.

Raw data were flagged for radio frequency interference (RFI), calibrated, and imaged following standard procedures with the Common Astronomy Software Application \textsc{casa}  \citep[v.5.1.2;][]{casa}. We first imaged the field with a Briggs robust parameter of zero to balance sensitivity and resolution, and reduce image side-lobes. We detected the radio counterpart of \src\, as a point source with peak flux density of $S_\nu$, of $4.38 \pm 0.05$\,mJy, where $\nu$ is the observing frequency (3\,GHz). We also measured an in-band spectral index, $\alpha$, of $-2 \pm 1$, where $S_\nu \propto \nu^{\alpha}$. 

\begin{figure*}[t]
\centering
\includegraphics[width=0.9\textwidth]{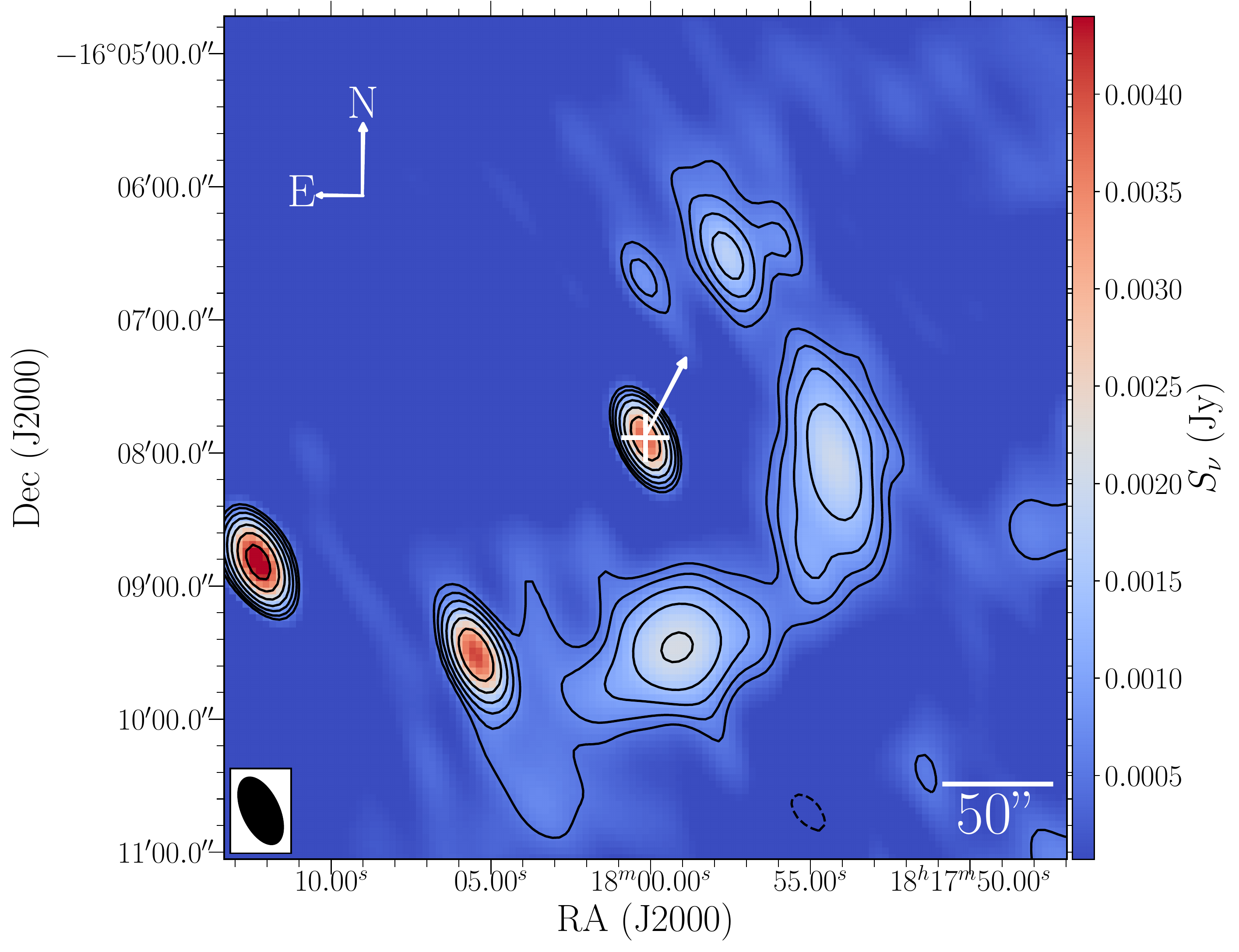}
\caption{3\,GHz VLA radio image of \src\ and the surrounding diffuse emission. This image was created using a natural weighting scheme to ensure that any diffuse emission was retained. The position of \src\ is marked by the white cross. The white arrow originating from the source position indicates its proper motion multiplied by a factor of 5000 \citep{ding22}. Contours are drawn at intervals of $\sqrt{2^n} \times \textrm{rms}$, where $n=5, 6, 7, 8,...$ and the image rms (near the source region) was 90\,$\mu$Jy/beam. The negative contour at the bottom of the figure is marked by dashed lines. 
The black-filled ellipse in the bottom-left corner represents the shape and size of the synthesized beam.
\src\ is clearly detected as a bright point source, which is surrounded by a half ring-like structure of diffuse emission.}
\label{fig:radio_emission} 
\end{figure*}

We also imaged the field with a Briggs robust parameter of two (corresponding to a natural weighting) to emphasise any diffuse emission in the field (Figure~\ref{fig:radio_emission}), although this did increase the image noise (to $\sim 0.09$\,mJy/beam). We detected a relatively bright (peaking at $\approx 2.2$\,mJy/beam) half-ring of diffuse emission located $\sim$90$^{\prime \prime}$ to the west of \src. The diffuse structure exhibits a radio spectral index between  $-1(\pm 1)$ and $-3(\pm 1)$. Unfortunately, from the radio image alone, we are unable to unambiguously connect this diffuse emission to \src, where, instead, the emission may be related to another source in the field (in particular the second bright source to the south-east of \src). However, taking into account the shape around \src\, we lean towards the scenario in which this diffuse radio emission is related to the magnetar. Further radio observations are planned/on-going to identify the nature and behaviour of this emission.

\section{Discussion}
\label{sec:disc}

We have presented the evolution of the X-ray spectral and timing properties of the magnetar \src\, following its first outburst with onset on 2020 March 12, as well as a VLA radio observation of the field. The X-ray monitoring campaign covered $\sim$19 months of the outburst decay, allowing us to characterize accurately the behavior of the source over a long time span.\\

\subsection{Long-term light curve modeling}

The 0.3--10\,keV luminosity reached a peak value of $\sim$9$\times$10$^{34}$\,\lum\ only a few minutes after the detection of the short burst that triggered \swift/BAT on 2020 March 12 \citep{evans20}. Then, it decreased down to $\sim$3$\times$10$^{33}$\,\lum\ after 575 days. To study the post-outburst luminosity decay, we modeled the temporal evolution of the 0.3--10\,keV luminosity with a phenomenological model consisting of an exponential function:
\begin{equation}
    L(t) = A \exp[-(t-t_0)/\tau]~,
\end{equation}
where $t_0$ is the epoch of the outburst onset fixed to MJD 58920.8866 (2020 March 12, 21:16:47 UTC; \citealt{evans20}), and the $e$-folding time $\tau$ can be interpreted as the decay timescale of the outburst. The fit resulted in $\tau = 153 \pm 1$ days for a reduced \rchisq = 0.8 for 59 d.o.f assuming an uncertainty of 20\% on all the nominal values of the luminosity. We can compare this result with that obtained by \citet{hu20}, who modeled the first $\sim$100 days of the luminosity temporal evolution using a double exponential function, giving $e$-folding timescales of $\tau_{1} = 9 \pm 2 $ and $\tau_{2} = 157 \pm 13 $ days. The latter reflects the decay trend on a longer timescale, and is fully consistent within the uncertainties with the value derived in this work using data covering the first $\sim$19 months of the outburst. We then integrated the best-fitting model over a time range spanning from the outburst onset (March 2020) to the last epoch of our observing campaign (October 2021), and estimated a total energy released in the outburst of $\approx10^{42}$\,erg. The reported results of the decay timescale and the released energy of the outburst of \src\, are in agreement with those derived by \citet{cotizelati18} for magnetars showing major outbursts (e.g., SGR\,1833--0832) and follow the correlation trend between these two quantities, implying that the decay pattern of this outburst is similar to that observed for other magnetar outbursts.\\

\begin{figure}
\resizebox{\hsize}{!}{\includegraphics[angle=0]{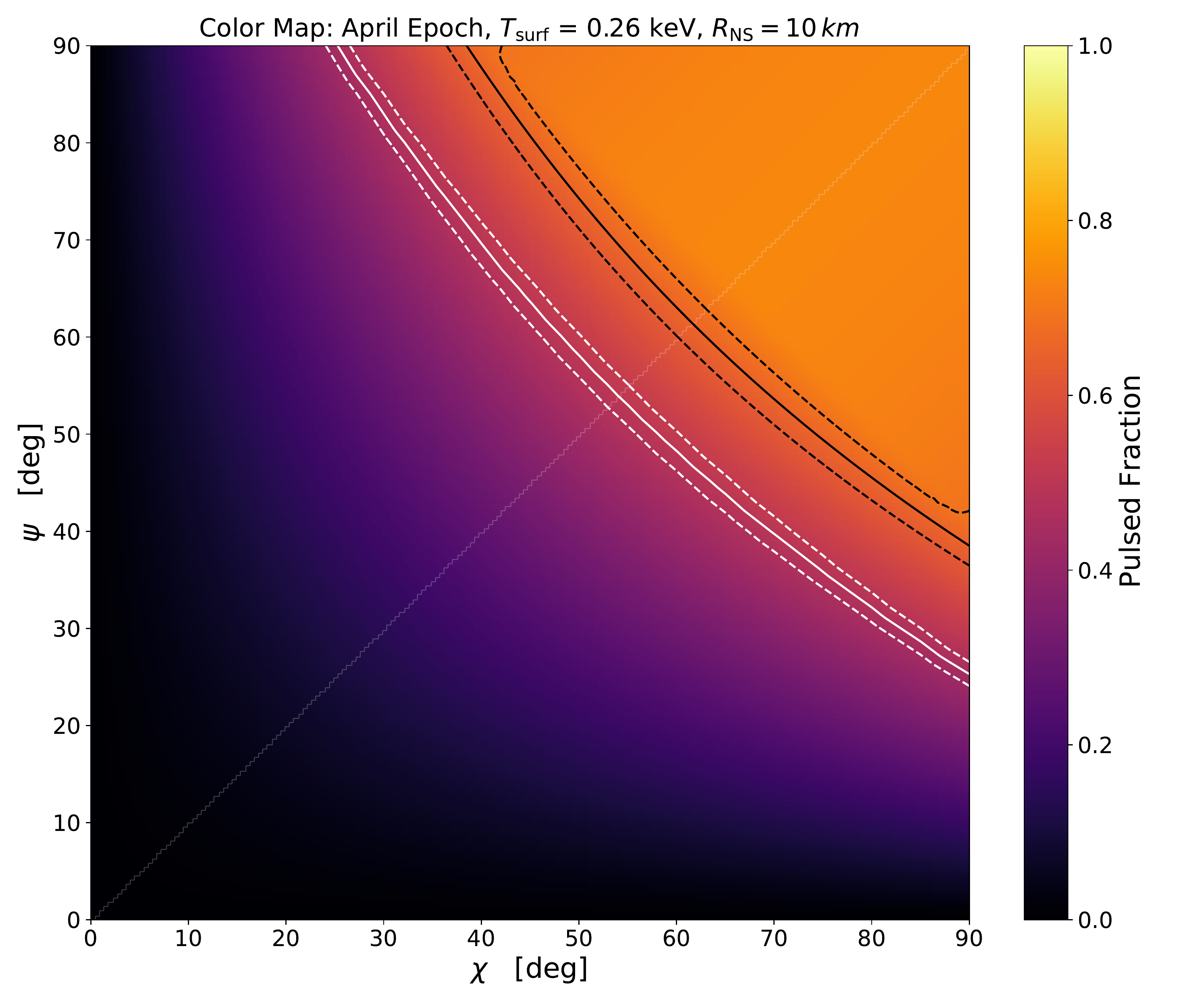}}
\caption{Constraints on the emission geometry of \src, based on the PF measured in April 2020. The color scale represents the 0.3--10\,keV PF at different angles calculated by employing a surface temperature of $kT_\textrm{star} = 0.26$\,keV. The black lines represent the measured value ($\textrm{PF} = 66\pm3 \%$). The white lines represent the same contours calculated considering only the flux from the hot-spot, neglecting the contribution from the remaining of the star.}
\label{fig:chipsiPF}
\end{figure}

\subsection{Timing analysis}

We attempted to derived a phase-connected timing solution from the \xmm\, and \nustar\, dataset, which covered a 7-months period from 2020 March to October. However, the sparsity of the observations lead to phase ambiguity and prevented us from identifying the correct timing model for the source. During the monitoring campaign, the spindown rate $\dot{\nu}$ of the source fluctuates but is overall increasing with time. Despite being poorly constrained, the spin evolution we observe is consistent with the timing results obtained by \cite{rajwade22} from the radio monitoring of the source after 2020 July, although we note a more rapid slowdown over the (earlier) time coverage of our X-ray data. At the reference epoch (MJD) 59022, we estimate a characteristic age $\tau_c\sim$510\,yr that is consistent with other measurements reported shortly following the onset of the outburst (e.g., \citealt{ccc+20,hu20}) but younger than the late-time radio measurements by \cite{rajwade22} (860\,yr) which is more likely to reflect the true spindown rate of the magnetar. Despite the torque variability of \src, it remains clear that this object is one of the youngest neutron stars in the Galaxy. However, a targeted monitoring campaign of the source during its quiescent state is needed to determine a more accurate secular spin-down rate.

\subsection{Constraining the emission geometry via pulsed-fraction modeling}

We constrain the emission geometry of \src\, namely the orientation of the hot spot with respect to the line of sight and the rotational axis, by comparing the PF observed in the \xmm\, data to a set of simulated PFs calculated using the approach described by \citet{Perna2001} and \citet{gotthelf10}.

We define a temperature map on the stellar surface, characterized by a uniform background temperature plus a single hot spot with a Gaussian temperature profile. The hot spot is oriented at an angle $\chi$ with respect to the star's rotational axis. After defining our line of sight as orientated at an angle $\psi$ with respect to the rotational axis, we calculate the observed phase-resolved spectra by integrating the local blackbody emission on the visible portion of the stellar surface. We take into account the gravitational light bending by approximating the ray-tracing function \citep{Pechenick1983, Page1995} with the formula derived by \citet{beloborodov02}. The interstellar medium absorption was also taken into account.

For each combination of ($\chi, \psi$) angles in the range $[0^\circ, 90^\circ]$, we integrate the phase-resolved spectra in the energy range 0.3--10\,keV to obtain a light curve, whose maximum and minimum values allow us to calculate the PF, according to Eq. (\ref{eq:pulsedfraction}).

We performed the analysis using the hot-spot parameters measured in April 2020 epoch ($kT_{BB} = 1.15 \, \textrm{keV}$ and $R_{BB} = 0.50\, \textrm{km}$) and October 2020 epoch ($kT_{BB} = 1.17 \, \textrm{keV}$ and $R_{BB} = 0.28\, \textrm{km}$). 
We used two different setups: a setup where we consider only the hot-spot contribution to the flux, neglecting the rest of the surface, and another setup, where we consider also the emission from the remaining surface, whose temperature is set as $kT_\textrm{star} = 0.26$\,keV (this latter setup describes a case where the contribution of the stellar background to the flux is maximal).
This value is an upper limit estimated by assuming uniform blackbody emission from the entire stellar surface (we have adopted a stellar radius of 10 km) and taking into account the effects of interstellar absorption (we fixed the absorption column density to \nh = 1.24 $\times 10^{23}$\,cm$^{-2}$): it is the one that gives an observed flux consistent with the count-rate upper limits derived using \xmm\ at pre-outburst epochs (0.008\,counts\,s$^{-1}$ at 3$\sigma$; see \citealt{esposito20}).

Figure\,\ref{fig:chipsiPF} shows the PF variation in the $\chi-\psi$ plane. The color scale represents the PF calculated using the hot-spots parameters derived from the April epoch with the second setup, where the contribution of the whole surface is considered. The black lines represent the constant PF contour equivalent to the measured value for this epoch, i.e., $\textrm{PF} = 66\pm 3 \%$. The white lines instead represent the same contour, but derived from the first setup, where only the hot spot emits. It is worth noticing that including the emission from the entire surface causes a shift of constant PF lines toward the upper-right corner. The reason is that this additional contribution acts in the direction of decreasing the PF, so that in order to keep the PF at the same value, we have to move towards higher values of $\chi$ and $\psi$. In this way, the hot-spot+surface and the only-hot-spot setups determine an upper and lower limit, respectively, in the $\chi-\psi$ plane. These limits constrain the geometry of the source. Since the result for the April epoch are more constraining than those for the October epoch, we report here only the former results. The geometrical constraints that we derive through modelling the rotation of the brightest hot spot cannot be directly compared with those usually derived via radio polarimetric observations \citep{Johnston2018}. During a magnetar outburst, the magnetosphere is very dynamic, and as a consequence the radio emission is highly variable in terms of intensity, pulse profile, polarization position angle (PA) on timescales ranging from hours to days (see \citealt{lower20}). For this reason, PA swings during magnetar outbursts are not a direct trace of the system geometry. On the other hand, the surface hot spot (which may or may not be located close to the magnetic pole) cannot move during the outburst and is a better proxy for the system geometry. For sources undergoing multiple outbursts, the properties of the heated spots may differ from event to event depending on the exact position and extension of the bundle that triggers that particular outburst. However, once formed, the hot spot cools down but does not move on the surface as the outburst decays.

\subsection{Spectral evolution of the source and diffuse emission}

We detected diffuse emission around \src\ in the \xmm/EPIC-pn observations, which confirms the result previously obtained by \cite{esposito20}. Extracting the surface brightness profiles, we found that the diffuse emission extends within 50\arcsec\,-- 110\arcsec\, (Figure \ref{diff_profile}, top panel). The spectra of this component are well described by a single blackbody model (Figure \ref{fig:diffusespec}). The best-fitting values show that the temperature of the diffuse emission $kT_{\rm diff}$ does not vary in time, while we see a clear decrease in the flux $F_{\rm X.diff}$ of $\approx$ 35\% between March and October 2020. Since an angular scale of 110\arcsec\, correspond to an extent of more than 8 light years at a distance of 4.8\,kpc, we can explain this variability in such a short time only by invoking a projection effect, such as in the case of a dust-scattering halo.

The study of the long-term spectral evolution of \src\, from EPIC-pn and FPMA observations showed that the X-ray spectrum is well described by a single blackbody, except for the epoch close to the outburst onset, where a power law was required to model the emission. To improve the sampling of the long-term magnetar flux evolution and trace the blackbody radius and temperature, we complemented these data sets with additional observations from the \swift/XRT telescope (see Table \ref{tab:observations1}).
Figure \ref{lcurve} shows the temporal evolution of the X-ray flux, the blackbody temperature and radius for \src. While the temperature remained constant around 1\,keV across the time span covered by the observations, the observed 0.3--10\,keV flux as well as the radius of the emitting region showed an exponentially decreasing trend. 

To compare the decay rate of the 0.3--10\,keV flux of the thermal and non-thermal components of \src\ in the early stages of the outburst, we evaluated the rate of the flux decrease separately for the two components over the period from March to April 2020. In the case of the \xmm+\nustar\ observations performed in April 2020, we added a power-law component in the spectral modeling, fixing the index at the value measured during the March epoch ($\Gamma = 1.04$). We found that, between the two epochs, the flux decay of the power-law component is about two times faster than that of the blackbody component.

\subsection{Point-like and diffuse radio emission}
\label{subsec:point-like diffuse radio emission}

Large radio flux variability was observed from \src\ during its outburst evolution, in line with what is typically seen in other radio-loud magnetars. The point-like continuum radio emission observed from \src\ in our VLA observations had a flux density of $4.38 \pm 0.05$\,mJy at 3\,GHz. Comparing this flux with the pulsed flux evolution reported by \cite{rajwade22} at 1.4\,GHz, the source continuum emission appears to be compatible with the reported pulsed flux assuming a flat spectrum between the two bands. The different bands, the large variability of the radio spectrum of this object, and the non-simultaneous flux measurements do not allow us to draw any strong conclusion about the presence of a non-pulsed continuum radio emission (i.e. a pulsar wind nebulae component).

Recent preliminary indication of a proper motion detected by VLBA \citep{ding22} hints to a motion of the source in the N-W direction, which would be toward the higher end of the observed radio ring-like structure. 

The presence of X-ray and radio diffuse emission on similar scales ($\sim90^{\prime\prime}$) might, in the first place, lead to a tentative association of these two emission regions. However, the variability we observe in the X-ray diffuse emission, its relatively soft X-ray spectrum, as well as the large $N_H$ we measure along the line of sight, lead us to interpret the X-ray diffuse emission as a dust scattering halo which is not expected to have radio counterparts. In addition, we stress that it is also possible that the diffuse radio emission is unrelated to \src.

If it is related to \src, the radio diffuse emission with its semi-circular and patchy appearance is similar to that observed in some supernova remnants. At a distance of 4.8\,kpc, the $\sim90^{\prime\prime}$ structure translates to a physical dimension of $\sim 2$\,pc. Assuming that this radio emission comes from the supernova remnant associated with the magnetar, we attempt to study its evolutionary status. The evolutionary path of a supernova remnant can be described in the radio domain by the $\Sigma-D$ diagram \cite[][Fig.\,3]{urosevic20}, where $\Sigma$ is the radio surface brightness and $D$ is the diameter.
We used the information obtained from our radio observation to assess the position of the remnant in this diagram and the associated evolutionary status.

The integrated flux density of the nebula is $\sim 3.6\times10^{-2}$~Jy at 3~GHz, from which we obtained a surface brightness at 1~GHz of $\Sigma \sim 1.1 \times 10^{-21}$~W~m~$^{-2}$~Hz~$^{-1}$~sr~$^{-1}$ assuming a spectral index $\alpha$ of 0.5 (corresponding to the mean radio spectral index of the observed Galactic supernova remnants; e.g. \citealt{dubner15}). Assuming a distance of $4.8$~kpc, this value would imply that the supernova remnant lies in the left corner of the $\Sigma-D$ diagram. Such a position is relative to a free expansion in an extremely-low density medium, which seems rather untenable (a similar scenario is discussed and rejected by \citealt{filipovic22} for the supernova remnant J0624--6948). On the other hand, if the source distance is 8.1\,kpc \cite[][]{ccc+20}, the supernova remnant would lie in the lower right corner of the diagram, meaning that it is in full Sedov phase. We also used the equipartition (eqp) calculator 
\cite[][\footnote{\url{http://poincare.matf.bg.ac.rs/~arbo/eqp/index.php?out=1##end}}]{urosevic20} to estimate the magnetic field strength by considering a distance of the remnant of 8.1\,kpc and a $\alpha = 0.5$. We obtained $B \sim 40$~$\mu$G, which is consistent (in terms of order of magnitude) with the value obtained for other well-studied supernova remnants in the same evolutionary phase \cite[see the case the middle-age Cygnus Loop supernova remnant][]{loru21}.

Further radio observations are planned to disentangle the spectrum of this diffuse radio emission, and possibly confirm its remnant nature (the corresponding results will be presented in a future paper).


\begin{acknowledgments}
AYI's work has been carried out within the framework of the doctoral program in Physics of the Universitat Autònoma de Barcelona. AYI, AB, NR, FCZ, EP, RS, SA, VG, CD and MR are supported by the H2020 ERC Consolidator Grant “MAGNESIA” under grant agreement No. 817661 (PI: Rea) and National Spanish grant PGC2018-095512-BI00. FCZ and VG are supported by Juan de la Cierva fellowships. AB acknowledge support from the Consejer\'ia de Econom\'ia, Conocimiento y Empleo del Gobierno de Canarias and the European Regional Development Fund (ERDF) under grant with reference ProID2021010132 ACCISI/FEDER, UE. TDR acknowledges financial contribution from the agreement ASI-INAF n.2017-14-H.0. MR acknowledges financial support from the Italian Ministry for Education, University and Research through grant 2017LJ39LM ‘UnIAM’ and the INAF ‘Main-streams’ funding grant (DP n.43/18).
SL acknowledges financial support from the Italian Ministry of University and Research - Project Proposal CIR01$\_$00010.
This work was also partially supported by the program Unidad de Excelencia Mar\'ia de Maeztu CEX2020-001058-M, and by the PHAROS COST Action (No. CA16214). 

\xmm\ is an ESA science mission with instruments and contributions directly funded by ESA member states and NASA.
\nustar\ is a project led by the California Institute of Technology, managed by the Jet Propulsion Laboratory and funded by NASA.
The National Radio Astronomy Observatory is a facility of the National Science Foundation operated under cooperative agreement by Associated Universities, Inc.

\facilities{\xmm, \nustar, \swift, \INT, VLA}
\software{CIAO (v6.29c; Fruscione et al. 2006), SAS (v19.1.0; Gabriel et al. 2004),  HEASoft pack-
age (v.6.29c; Nasa High Energy Astrophysics Science
Archive Research Center (Heasarc) 2014),   NUSTARDAS (v.2.1.1) and CALDB (v.20211202),
TEMPO timing software (Nice et al. 2015), CASA (v5.1.2; THE CASA TEAM et al. 2022), Xspec (v12.11.1; Arnaud 1996)}

\end{acknowledgments}


\bibliography{biblio}{}
\bibliographystyle{aasjournal}



\appendix
\restartappendixnumbering 

\section{Observation log of \srclong}
\startlongtable
\begin{deluxetable*}{cccccc}
\tablecaption{Observation log of \src, including the observations analysed by \citet{esposito20} above the double-horizontal solid lines.
\label{tab:observations1}}
\tabletypesize{\scriptsize}
\tablecolumns{5}
\tablewidth{0pt}
\tablehead{
\colhead{Instrument\tablenotemark{a}} &
\colhead{Obs.ID} &
\colhead{Start} &
\colhead{Stop} & 
\colhead{Exposure} & \colhead{Count rate\tablenotemark{b}}\\
\colhead{} &
\colhead{} & 
\multicolumn2c{YYYY-MM-DD hh:mm:ss (TT)} &
\colhead{(ks)} & \colhead{(counts s$^{-1}$)}
}
\startdata
\swift/XRT (PC) & 00960986000 & 2020-03-12 21:18:22 & 2020-03-12 21:36:48 & 1.1 & 0.15 $\pm$ 0.01 \\
\swift/XRT (PC) & 00960986001 & 2020-03-12 22:57:45 & 2020-03-13 05:13:02 & 4.9 &  0.14 $\pm$ 0.01 \\
\swift/XRT (WT) & 00960986002 & 2020-03-13 20:47:55 & 2020-03-13 21:21:15 & 2.0 &  0.16 $\pm$ 0.01 \\
\swift/XRT (PC) & 00960986003 & 2020-03-15 00:10:37 & 2020-03-15 03:36:52 & 1.5 &  0.14 $\pm$ 0.01 \\
\nustar/FPMA & 80402308002 & 2020-03-15 03:58:21 & 2020-03-15 15:58:03 & 22.2 & $0.443\pm0.005$  \\
\emph{XMM}/EPIC-pn (LW)  & 0823591801 & 2020-03-15 07:57:47 & 2020-03-15 14:41:12  & 22.1 & $1.45\pm0.01$ \\
\swift/XRT (WT) & 00960986004 & 2020-03-19 09:33:11 & 2020-03-19 11:16:56 & 1.7 & 0.19 $\pm$ 0.02 \\
\swift/XRT (WT) & 00960986005 & 2020-03-20 04:34:19 & 2020-03-20 04:49:56 & 1.8 & 0.20 $\pm$ 0.01 \\
\swift/XRT (WT) & 00960986006 & 2020-03-22 02:35:21 & 2020-03-22 03:01:56 & 1.6 & 0.16 $\pm$ 0.01 \\
\swift/XRT (WT) & 00960986007 & 2020-03-24 05:51:38 & 2020-03-24 09:02:56 & 1.2 & 0.13 $\pm$ 0.01 \\
\swift/XRT (WT) & 00960986008 & 2020-03-26 05:40:29 & 2020-03-26 23:20:56 & 1.1 & 0.19 $\pm$ 0.01 \\
\swift/XRT (WT) & 00960986009 & 2020-03-28 03:40:53 & 2020-03-28 18:07:56 & 1.2 & 0.18 $\pm$ 0.02 \\
\swift/XRT (WT) & 00960986010 & 2020-03-29 16:25:13 & 2020-03-30 21:03:56 & 1.3 & 0.16 $\pm$ 0.01 \\
\swift/XRT (WT) & 00960986011 & 2020-04-01 19:17:34 & 2020-04-01 19:25:56 & 0.5 & 0.17 $\pm$ 0.02 \\
\hline
\hline
\nustar/FPMA  & 80402308004 & 2020-04-04 02:01:09 & 2020-04-05 13:36:09 & 59.1 & 0.339 $\pm$ 0.002 \\
\emph{XMM}/EPIC-pn (FF) & 0823593901 & 2020-04-04 03:44:15 & 2020-04-04 13:32:52 & 33.4 & 1.08 $\pm$ 0.01 \\
\swift/XRT (WT) & 00089033001 & 2020-04-05 05:51:22 & 2020-04-05 06:18:56 & 1.6 & 0.13 $\pm$ 0.01 \\
\swift/XRT (WT) & 00960986012 & 2020-04-17 11:22:40 & 2020-04-18 17:58:56 & 2.1 & 0.22 $\pm$ 0.01 \\
\swift/XRT (WT) & 00960986013 & 2020-04-21 01:15:46 & 2020-04-21 12:42:55 & 1.3 & 0.17 $\pm$ 0.01 \\
\swift/XRT (WT) & 00960986014 & 2020-05-02 07:58:09 & 2020-05-02 12:50:56 & 1.3 & 0.13 $\pm$ 0.02 \\
\nustar/FPMA & 80402308006 & 2020-05-02 20:56:09 & 2020-05-03 20:26:09 & 42.2 &  0.277 $\pm$ 0.003\\
\swift/XRT (PC) & 00089033002 & 2020-05-02 22:22:26 & 2020-05-02 22:49:53 & 1.6 & 0.08 $\pm$ 0.01 \\
\swift/XRT (PC) & 00969823991$^{c1}$ & 2020-05-06 17:38:19 & 2020-05-06 17:48:46 & 0.6 & 0.08 $\pm$ 0.01 \\
\swift/XRT (PC) & 00969823001$^{c1}$ & 2020-05-06 18:43:12 & 2020-05-06 20:42:42 & 1.9 & 0.09 $\pm$ 0.01 \\
\swift/XRT (WT) & 00969823002 & 2020-05-13 05:39:09 & 2020-05-13 10:16:56 & 1.3 & 0.13 $\pm$ 0.01 \\ 
\swift/XRT (WT) & 00969823003 & 2020-05-15 05:19:42 & 2020-05-15 20:02:56 & 1.6 & 0.13 $\pm$ 0.01 \\
\swift/XRT (PC) & 00972614991 & 2020-05-16 15:04:54 & 2020-05-16 16:27:25 & 1.7 & 0.08 $\pm$ 0.01 \\ 
\swift/XRT (WT) & 00969823004 & 2020-05-21 20:38:46 & 2020-05-21 22:24:56 & 1.9 & 0.10 $\pm$ 0.01 \\
\swift/XRT (WT) & 00969823005 & 2020-05-28 00:46:46 & 2020-05-28 20:09:56 & 2.5 & 0.14 $\pm$ 0.01 \\
\swift/XRT (WT) & 00969823006 & 2020-06-04 08:14:35 & 2020-06-04 18:08:56 & 2.7 & 0.12 $\pm$ 0.01 \\
\swift/XRT (WT) & 00969823007$^{c2}$ & 2020-06-12 01:11:04 & 2020-06-13 07:21:56 & 1.5 & 0.13 $\pm$ 0.01 \\
\swift/XRT (WT) & 00969823008$^{c2}$ & 2020-06-15 08:53:23 & 2020-06-15 12:12:55 & 0.4 & 0.12 $\pm$ 0.02 \\
\swift/XRT (WT) & 00969823009$^{c2}$ & 2020-06-18 19:16:44 & 2020-06-18 19:21:56 & 0.3 & 0.10 $\pm$ 0.02 \\
\swift/XRT (WT) & 00969823010 & 2020-06-21 20:47:49 & 2020-06-22 21:07:56 & 3.1 & 0.05 $\pm$ 0.01 \\
\swift/XRT (WT) & 00969823011 & 2020-06-25 05:54:07 & 2020-06-25 23:46:56 & 3.1 & 0.07 $\pm$ 0.01 \\
\nustar/FPMA    & 80402308008 & 2020-06-30 02:46:09 & 2020-06-30 15:16:09 & 23.5 & 0.129 $\pm$ 0.002 \\
\nustar/FPMA   & 80402308010 & 2020-07-01 19:01:09 & 2020-07-02 01:01:09 & 12.3 &  0.161 $\pm$ 0.003 \\
\swift/XRT (PC) & 00089033003 & 2020-07-01 19:49:42 & 2020-07-01 20:12:54 & 1.4 & 0.041 $\pm$ 0.006 \\
\swift/XRT (PC) & 00980513991 & 2020-07-02 07:11:35 & 2020-07-02T08:34:14 & 1.7 & 0.061 $\pm$ 0.006 \\
\swift/XRT (WT) & 00969823012 & 2020-07-02 08:49:43 & 2020-07-02 16:22:56 & 1.7 & 0.059 $\pm$ 0.006 \\
\swift/XRT (PC) & 00089033004 & 2020-07-10 01:31:26 & 2020-07-10 08:10:54 & 3.6 & 0.059 $\pm$ 0.004 \\
\swift/XRT (PC) & 00089033005 & 2020-07-26 12:23:40 & 2020-07-26 18:58:52 & 3.8 & 0.050 $\pm$ 0.003 \\
\swift/XRT (PC) & 00089033006$^{c3}$ & 2020-08-07 03:12:00 & 2020-08-07 08:10:52 & 2.8 & 0.057 $\pm$ 0.005 \\
\swift/XRT (PC) & 00089033007$^{c3}$ & 2020-08-12 17:14:14 & 2020-08-12 17:33:52 & 1.2 & 0.031 $\pm$ 0.005 \\
\swift/XRT (PC) & 00089033008 & 2020-08-21 00:20:13 & 2020-08-21 14:47:27 & 3.0 & 0.044 $\pm$ 0.004 \\
\swift/XRT (PC) & 00089033009$^{c4}$ & 2020-09-04 07:12:43 & 2020-09-04 10:36:53 & 1.4 & 0.041 $\pm$ 0.006 \\
\nustar/FPMA & 80402308012$^{d}$ & 2020-09-07 00:41:09 & 2020-09-08 00:11:09 & 39.3  & 0.097 $\pm$ 0.001    \\
\emph{XMM}/EPIC-pn (FF) & 0823594001 & 2020-09-07 15:12:48 & 2020-09-07 22:38:50 & 26.7 & 0.429 $\pm$ 0.005\\ 
\swift/XRT (PC) & 00089033010$^{c4}$ & 2020-09-07 17:58:15 & 2020-09-07 19:30:52 & 2.0 & 0.043 $\pm$ 0.005 \\
\swift/XRT (PC) & 00089033011 & 2020-09-11 17:17:33 & 2020-09-12 04:34:52 & 4.8 & 0.044 $\pm$ 0.003 \\
\swift/XRT (PC) & 00089033012 & 2020-09-26 00:03:15 & 2020-09-26 20:39:53 & 4.5 & 0.040 $\pm$ 0.003 \\
\emph{XMM}/EPIC-pn (FF) & 0823594201 & 2020-10-08 10:24:17  & 2020-10-09 00:49:31  & 49.4 &  0.288 $\pm$ 0.004 \\
\swift/XRT (PC) & 00089033013 & 2020-10-08 14:47:04 & 2020-10-09 20:58:54 & 4.8 & 0.036 $\pm$ 0.003 \\ 
\swift/XRT (PC) & 03110882001 & 2020-10-19 08:54:19 & 2020-10-19 23:37:52 & 4.7 & 0.037 $\pm$ 0.003 \\
\swift/XRT (PC) & 00089033014 & 2020-10-25 09:43:50 & 2020-10-26 23:59:52 & 4.2 & 0.038 $\pm$ 0.003 \\
\swift/XRT (PC) & 00089033015 & 2020-11-06 06:55:28 & 2020-11-06 16:48:54 & 4.2 & 0.035 $\pm$ 0.003 \\
\swift/XRT (PC) & 00013996001 & 2021-02-15 09:24:24 & 2021-02-15 22:35:52 & 4.5 & 0.018 $\pm$ 0.002 \\
\swift/XRT (PC) & 00013996002 & 2021-03-01 00:16:23 & 2021-03-01 21:05:53 & 3.8 & 0.013 $\pm$ 0.002 \\
\swift/XRT (PC) & 00013996003 & 2021-03-15 14:26:25 & 2021-03-15 17:58:53 & 4.6 & 0.009 $\pm$ 0.002 \\
\swift/XRT (PC) & 00013996004 & 2021-03-29 01:52:19 & 2021-03-29 10:10:54 & 4.5 & 0.009 $\pm$ 0.001 \\
\swift/XRT (PC) & 00013996005 & 2021-04-12 03:40:52 & 2021-04-12 18:18:52 & 4.3 & 0.011 $\pm$ 0.002 \\
\swift/XRT (PC) & 00013996006 & 2021-04-26 02:28:48 & 2021-04-26 23:20:53 & 4.2 & 0.010 $\pm$ 0.002 \\
\swift/XRT (PC) & 00013996007 & 2021-05-10 02:30:39 & 2021-05-10 20:22:55 & 4.5 & 0.008 $\pm$ 0.002 \\
\swift/XRT (PC) & 00013996008$^{c5}$ & 2021-05-24 02:30:31 & 2021-05-24 10:38:17 & 2.3 & 0.008 $\pm$ 0.002 \\
\swift/XRT (PC) & 00013996009$^{c5}$ & 2021-05-29 19:27:09 & 2021-05-29 22:51:54 & 2.4 & 0.005 $\pm$ 0.002 \\
\swift/XRT (PC) & 00013996010$^{c5}$ & 2021-06-06 03:11:41 & 2021-06-07 22:04:51 & 1.6 & 0.010 $\pm$ 0.002 \\
\swift/XRT (PC) & 00013996011$^{c6}$ & 2021-06-26 18:32:43 & 2021-06-26 18:41:52 & 0.5 & 0.008 $\pm$ 0.004 \\
\swift/XRT (PC) & 00013996012$^{c6}$ & 2021-06-28 00:48:50 & 2021-06-28 18:02:52 & 1.0 & 0.007 $\pm$ 0.003 \\
\swift/XRT (PC) & 00013996013$^{c6}$ & 2021-06-29 06:56:55 & 2021-06-29 21:32:54 & 2.1 & 0.007 $\pm$ 0.002 \\
\swift/XRT (PC) & 00013996014$^{c6}$ & 2021-06-30 02:07:20 & 2021-06-30 14:40:52 & 2.7 & 0.008 $\pm$ 0.002 \\
\swift/XRT (PC) & 00013996015$^{c7}$ & 2021-07-05 11:13:57 & 2021-07-06 17:19:06 & 1.0 & 0.007 $\pm$ 0.003 \\ 
\swift/XRT (PC) & 00013996016$^{c7}$ & 2021-07-08 01:13:22 & 2021-07-08 12:38:52 & 0.6 & 0.002 $\pm$ 0.002 \\
\swift/XRT (PC) & 00013996017$^{c7}$ & 2021-07-13 06:45:26 & 2021-07-13 18:11:53 & 2.6 & 0.005 $\pm$ 0.002 \\ 
\swift/XRT (PC) & 00013996018$^{c7}$ & 2021-07-18 01:52:35 & 2021-07-18 23:59:52 & 0.3 & 0.005 $\pm$ 0.001 \\
\swift/XRT (PC) & 00013996019$^{c8}$ & 2021-08-08 10:46:46 & 2021-08-08 12:26:54 & 0.6 & 0.010 $\pm$ 0.005 \\
\swift/XRT (PC) & 00013996020$^{c8}$ & 2021-08-10 08:34:29 & 2021-08-10 08:45:53 & 0.7 & 0.010 $\pm$ 0.004 \\
\swift/XRT (PC) & 00013996021$^{c9}$ & 2021-09-24 02:13:29 & 2021-09-24 23:07:52 & 4.1 & 0.006$\pm$0.001 \\ 
\swift/XRT (PC) & 00013996022$^{c9}$ & 2021-10-24 02:27:38 & 2021-10-25 15:15:53 & 4.4 & 0.004$\pm$0.001 \\
\enddata
\tablenotetext{a}{The instrumental setup is indicated in parentheses: PC = photon counting, WT = windowed timing, LW = large window, and FF = full frame.}
\tablenotetext{b}{The count rate is in the 0.3--10\,keV energy range, except for \nustar\ (3--10\,keV).}
\tablenotetext{c}{Observations with the same superscripts were merged for the spectral analysis.}
\tablenotetext{d}{Data collected with FPMB were not included in the analysis as they are heavily affected by stray light contamination.}
\end{deluxetable*}

\section{\nustar\ bursts: fluence and duration}

In Table\,\ref{tab:bursts}, we report the properties for the bursts detected in the \nustar\ light curves. In the table, the epochs are referred to the Solar system barycenter, the fluence refers to the 3--79\,keV range and the duration has to be considered as an approximate value. We estimated it by summing the 15.625-ms time bins showing enhanced emission for the structured bursts, and by setting it equal to the coarser time resolution at which the burst is detected in all the other cases. Therefore, it has to be considered as an approximate value. Except for burst 1 on 2020 May 3 (125\,ms), all the remaining bursts have a duration of 62.5\,ms.

\begin{table}
\centering
\caption{Log of X-ray bursts detected in the \nustar\ light curves. 
\label{tab:bursts}}
\begin{tabular}{cccc}
\hline
\hline
Obs.ID & \multicolumn{2}{c}{Burst epoch} & Fluence \\
 & \multicolumn{2}{c}{YYYY-MM-DD hh:mm:ss (TDB)} & (counts) \\ 
\hline
80402308002 & 2020-03-15 & 05:25:59 & 10 \\
            &            & 05:45:51 & 8 \\
80402308002 & 2020-04-05 & 04:12:22 & 7 \\
80402308002 & 2020-05-03 & 18:05:07 & 22 \\
\hline
\end{tabular}
\end{table}

\end{document}